\RequirePackage{lineno}
\documentclass[floatfix,twocolumn,prd,showpacs,preprintnumbers,amsmath,nofootinbib,amssymb,noeprint,superscriptaddress]{revtex4-1}

\usepackage[normalem]{ulem}

\usepackage{mathtools, cuted}
\usepackage{lipsum}
\usepackage[utf8]{inputenc}
\usepackage{graphicx}
\usepackage{dsfont}
\usepackage{mathrsfs}
\usepackage{bm}
\usepackage{color}
\usepackage[colorlinks,citecolor=blue,urlcolor=blue,linkcolor=black]{hyperref}
\usepackage{url}
\usepackage{braket}
\usepackage{placeins}
\usepackage{multirow}
\usepackage{hyperref}
\usepackage{float}
\usepackage{booktabs}
\usepackage{float}
\usepackage{comment}

\newcommand{\err}[3]{${#1}^{+ #2}_{- #3}$}

\def\tauf{\tau_{\mathrm{F}}}
\def\msbar{\overline{\mathrm{MS}}}

\usepackage[nameinlink,capitalize]{cleveref}
\usepackage{diagbox}

\begin{document}
\title{Shear and bulk viscosities of the gluon plasma across the transition temperature from lattice QCD}


\author{Heng-Tong Ding}
\affiliation{Key Laboratory of Quark \& Lepton Physics (MOE) and Institute of Particle Physics, Central China Normal University, Wuhan 430079, China}

\author{Hai-Tao Shu}
\email{hai-tao.shu@ccnu.edu.cn}
\affiliation{Key Laboratory of Quark \& Lepton Physics (MOE) and Institute of Particle Physics, Central China Normal University, Wuhan 430079, China}

\author{Cheng Zhang}
\email{chengzhang@mails.ccnu.edu.cn}
\affiliation{Key Laboratory of Quark \& Lepton Physics (MOE) and Institute of Particle Physics, Central China Normal University, Wuhan 430079, China}

\begin{abstract}
We investigate the temperature dependence of the shear viscosity ($\eta$) and bulk viscosity ($\zeta$) of the gluon plasma using lattice QCD over the range  0.76--2.25$\,T_c$, extending from below the transition temperature $T_c$ across the transition region and into the deconfined phase. At each temperature, we employ three large, fine lattices, which enables controlled continuum extrapolations of the energy-momentum tensor correlators. Using gradient flow together with a recently developed blocking technique, we achieve percent-level precision for these correlators, providing strong constraints for a model-based spectral analysis. Since the inversion to real-time information is intrinsically ill posed, we extract viscosities by fitting spectral functions whose ultraviolet behavior is matched to the best available perturbative result, while the infrared region is described by a Lorentzian transport peak. The dominant modeling uncertainty associated with the transport peak width is bracketed by varying it over a physically motivated range set by thermal scales.
We find that the shear-viscosity-to-entropy-density ratio, $\eta/s$, exhibits a minimum near the transition temperature $T_c$ and increases for $T>T_c$, whereas the bulk-viscosity-to-entropy-density ratio, $\zeta/s$, decreases monotonically over the entire temperature range studied.

\end{abstract}

\maketitle

\section{Introduction}
\label{sec:intro}

Shear viscosity $\eta$ quantifies a fluid’s resistance to deformation under shear stress, while bulk viscosity $\zeta$ measures its resistance to uniform compression or expansion. Both viscosities play a central role in describing the collective dynamical behavior of the quark-gluon plasma (QGP) created in heavy-ion collisions. Vanishing values of these viscosities correspond to ideal (perfect) fluid behavior. As fundamental parameters characterizing the transport properties of the QGP, viscosities have, therefore, attracted substantial experimental and theoretical interests \cite{Teaney:2003kp,Kovtun:2004de,Lacey:2006bc,Csernai:2006zz,Meyer:2007ic,Meyer:2007dy,Kharzeev:2007wb,Marty:2013ita,Christiansen:2014ypa,Astrakhantsev:2017nrs,Borsanyi:2018srz,Ghiglieri:2018dib,Astrakhantsev:2018oue,Mykhaylova:2019wci,Itou:2020azb,Mykhaylova:2020pfk,Altenkort:2022yhb}. In particular, phenomenological studies~\cite{PHENIX:2006iih,Drescher:2007cd,Dusling:2007gi,Romatschke:2007mq,Xu:2007jv,Luzum:2008cw,Chaudhuri:2009uk,ALICE:2011ab,Bernhard:2019bmu,Shen:2023awv} as well as analyses combining viscous hydrodynamics with a microscopic transport model~\cite{Song:2010mg} typically favor small viscosities, with the latter yielding $\eta/s \in [1/(4\pi),2.5/(4\pi)]$. The lower end of this range coincides with the lower bound of the AdS/CFT result $\eta/s \ge 1/(4\pi)$, which is obtained for a strongly coupled deconfined plasma~\cite{Kovtun:2004de}, while the upper end is close to the next-to-leading-order (NLO) weak-coupling QCD estimate $\eta/s \sim 0.25$ at temperature near $T_c$~\cite{Ghiglieri:2018dib}. However, the substantial discrepancy between the NLO prediction and the leading-order (LO) result, $\eta/s \sim 0.9$~\cite{Arnold:2003zc}, indicates slow convergence of the perturbative calculations. This motivates a genuinely nonperturbative determination of the shear viscosity from first principles using lattice QCD. In contrast,
perturbative calculations of the bulk viscosity are technically more involved and, so far, have only been carried out at leading order~\cite{Arnold:2006fz}---higher-order corrections are not yet available, making lattice input particularly valuable.

In fact, lattice computations of the shear and bulk viscosities have already been carried out in several studies for pure gauge theory~\cite{Meyer:2007dy,Meyer:2007ic,Astrakhantsev:2017nrs,Astrakhantsev:2018oue,Borsanyi:2018srz}, primarily using the multilevel (ML) algorithm~\cite{Luscher:2001up,Ce:2016ajy,Giusti:2022xdh} to suppress the severe ultraviolet noise in the relevant Euclidean energy-momentum tensor (EMT) correlators. However, the ML algorithm is not straightforwardly applicable to full QCD with dynamical quarks\footnote{A recent extension of the multilevel algorithm to full QCD has been reported in Ref.~\cite{Barca:2025dca}, but it has not yet been used for the extraction of transport coefficients.}, which calls for a more general strategy.  The gradient flow (GF) method provides such an approach and is among the most practical options currently available for full QCD calculations~\cite{Narayanan:2006rf,Luscher:2010iy,Luscher:2013cpa}. GF has already been successfully used in a wide range of transport-related lattice studies~\cite{Kitazawa:2017qab,Altenkort:2020fgs,Altenkort:2020axj,Altenkort:2022yhb,Altenkort:2024spl,Itou:2020azb}, including a first attempt at extracting the shear viscosity in Ref.~\cite{Itou:2020azb}. That study demonstrated, however, that the use of GF alone does not deliver EMT correlators with sufficient precision to tightly constrain the ensuing spectral analysis. 

To remedy the insufficient precision, a recently developed blocking method~\cite{Altenkort:2021jbk} was incorporated as a complementary technique. It further improves the signal quality by a factor of 3--7, without increasing the computational cost. This combined GF and blocking strategy was first employed in a viscosity calculation at a single temperature, 1.5$\,T_c$~\cite{Altenkort:2022yhb}. 
There, the modeling systematics of the spectral reconstruction were explored by considering different interpolations between the infrared and ultraviolet regimes. In the ultraviolet region, the most accurate perturbative input currently available, namely the NLO spectral function, was used, while the infrared part was taken either with a linear behavior in frequency or with a Lorentzian peak whose width was varied within a physically motivated range set by thermal scales. This provided a systematic way to bracket the dominant modeling uncertainty and to extract viscosities with quantified systematics.

In this work, we extend the analysis of Ref.~\cite{Altenkort:2022yhb}, to which one of the present authors contributed, to a wide temperature range, 0.76--2.25$\,T_c$, covering temperatures below $T_c$, across the transition region, and deep into the deconfined phase. This exceeds the coverage of earlier lattice studies based on the ML algorithm~\cite{Astrakhantsev:2017nrs,Astrakhantsev:2018oue}, which were restricted to 0.9--1.5$\,T_c$. Moreover, unlike Refs.~\cite{Astrakhantsev:2017nrs,Astrakhantsev:2018oue}, which used small, relatively coarse lattices and did not perform continuum extrapolations, in this study we employ three lattices at each temperature that are both large and fine, with maximal spatial extent $L = 3.31~\mathrm{fm}$ and minimal lattice spacing $a = 0.01164~\mathrm{fm}$. This setup allows controlled continuum extrapolations throughout the full temperature range. The resulting EMT correlators reach percent-level precision, providing stringent constraints for a systematic-controlled extraction of the shear and bulk viscosities via spectral reconstruction.

The remainder of this paper is organized as follows. Section~\ref{sec:gradflow} introduces the definition of the EMT within the gradient-flow framework and outlines our strategy for extracting viscosities.  Section~\ref{sec:comp-emt} describes the lattice setup and the semi-nonperturbative renormalization of the EMT correlators. Continuum and flow-time extrapolations are presented next, followed by the spectral analysis in Sec.~\ref{sec:spectrum}.  Our main results and conclusions are summarized in Sec.~\ref{sec:conclusion}.  Readers primarily interested in the physics results rather than computational details may proceed directly to Sec.~\ref{sec:conclusion}.

\section{Viscosities in the framework of gradient flow}
\label{sec:gradflow}
Shear and bulk viscosities can be determined from the corresponding spectral function $\rho(\omega,T)$ via the Kubo formulae 
\begin{equation}
\begin{split}
    \eta(T) & =\lim_{\omega\rightarrow  0} \frac{\rho_{\rm{shear}}(\omega,T)}{\omega}, \\
    \zeta(T) & = \frac{1}{9} \lim_{\omega \to 0} \frac{\rho_{\rm{bulk}}(\omega,T)}{\omega},
\end{split}
\label{eq:kubo}
\end{equation}
as a consequence of the linear response theory~\cite{Jeon:1994if}. The spectral function can be extracted from the Euclidean correlation function $G(\tau)$ through the following integral transform~\cite{Yagi:2005yb}:
\begin{align}
\label{eq:correlator-spf}
G(\tau)=\int_0^{\infty}\frac{\mathrm{d}\omega}{\pi}\frac{\cosh[\omega(1/2T-\tau)]}{\sinh(\omega/2T)}\rho(\omega,T)\,,
\end{align}
where periodic conditions are imposed in the imaginary time $\tau$ direction. This universal formula allows for the determination of transport coefficients using lattice QCD.
To compute shear and bulk viscosity, we evaluate the correlation functions of the energy-momentum tensor, denoted as $T_{\mu\nu}(x)$, in the shear and bulk channels, respectively. Specifically we measure the correlation function of the traceless part and the trace anomaly of $T_{\mu\nu}(x)$,
\begin{align}
    \begin{split}
    &G_{\rm{shear}}(\tau)=\frac{1}{10} \int \mathrm{d}^3x\ \left\langle \pi_{ij}(0,\vec{0}) \: \pi_{ij}(\tau,\vec x)
     \right\rangle,\\
    &G_{\rm{bulk}}(\tau)=\int \mathrm{d}^3x\ \left\langle T_{\mu\mu}(0,\vec{0}) \: T_{\mu\mu}(\tau,\vec{x})\right\rangle,
    \label{eq:Gshearbulk}
    \end{split}
\end{align}
where $\pi_{ij} = T_{ij} - \frac{1}{3} \delta_{ij} T_{kk}$. \footnote{Here, we define the shear channel correlation functions using $\pi_{ij}$ rather than the commonly used $T_{12}$ for two reasons: (i) at vanishing momentum, which is our case, three spatial directions are equivalent so one can choose any $T_{ij}$, and (ii) the  rotational invariance~\cite{Meyer:2009vj} provides a one-to-one correspondence between the diagonal part and the off-diagonal part of $\pi_{ij}$. Therefore, using this definition could increase the statistics by a factor of 6. The number 10 is a normalization to account for the difference between $T_{12}$ and $\pi_{ij}$.}

A proper discretization and renormalization of the EMT operator on the lattice is nontrivial due to the breaking of continuous translational symmetry, see, e.g., ~\cite{Giusti:2015daa,DallaBrida:2020gux} for detailed discussions. Fortunately, gradient flow provides a feasible framework in which the different components of $\pi_{ij}$ renormalize in the same manner. In this approach, the EMT can be written in terms of a traceless operator $U_{\mu\nu}$ and a scalar operator $E$~\cite{Suzuki:2013gza},
\begin{equation}
    \label{EMT_flow}
    T_{\mu\nu}(x, \tauf)=c_1(\tauf) U_{\mu\nu}(x, \tauf)+4c_2(\tauf)\delta_{\mu\nu}E(x, \tauf)\, .
\end{equation}
Here, $c_1$ and $c_2$ are the renormalization constants that are known perturbatively up to two-loop and three-loop order, respectively~\cite{Suzuki:2021tlr}. In this work, we determine $c_1$ and $c_2$ semi-nonperturbatively---details are given in Sec.~\ref{sec:sec_renormalization}. The operators $U_{\mu\nu}$ and $E$ are constructed from the flowed field strength tensor $G^a_{\mu\nu}(x,\tauf)$ as
\begin{equation}
    \label{eq:E_U}
    \begin{split}
    U_{\mu\nu}(x,\tauf) &=G^a_{\mu\rho}(x,\tauf)G^a_{\nu\rho}(x,\tauf)-\delta_{\mu\nu}E(x, \tauf)\,,\\
    E(x, \tauf) &=\frac{1}{4}G^a_{\rho\sigma}(x,\tauf)G^a_{\rho\sigma}(x,\tauf)\, ,
    \end{split}
\end{equation}
where $G^a_{\mu\nu}(x,\tauf)$ is discretized using the standard clover definition evaluated on the flowed gauge fields $B_\mu(x,\tauf)$. These fields evolve from the original, unflowed gauge fields $A_\mu(x)$ along a fictitious fifth dimension---the flow time \(\tauf\)---according to~\cite{Luscher:2010iy,Luscher:2013cpa}
\begin{align}
    \label{eq:flow1}
    &B_\nu(x,\tauf=0)  = A_\nu(x) \,,
    \nonumber \\
    & \dot{B}_{\mu} = D_{\nu}G_{\nu\mu}\,,
\end{align}
where the dot denotes derivative with respect to the flow time and 
\begin{align}
    \label{eq:flow2}
    D_{\mu} &= \partial_{\mu}+[B_{\mu},\cdot] \,, \nonumber \\
   G_{\mu\nu}  &= \partial_{\mu}B_{\nu}-\partial_{\nu}B_{\mu}+[B_{\mu},B_{\nu}]\,,
\end{align}
are the flowed covariant derivative and field strength tensor, respectively.

At finite flow time, the strong UV fluctuations of the gauge fields are suppressed by flow smearing effects~\cite{Luscher:2010iy}. This suppression enhances the signal of gauge field-composite operators and induces a predictable flow-time dependence within an appropriate flow-time window \cite{Suzuki:2021tlr}. The precise correlators obtained at finite $\tauf$ are then extrapolated to the $\tauf\rightarrow 0$ limit to recover the corresponding renormalized operators. This procedure is discussed in detail in Sec.~\ref{sec:extrapolate}.

The high-precision correlators on the left-hand side of the convolution equation, Eq.~(\ref{eq:correlator-spf}), are used to reconstruct the spectral function on the right-hand side. This reconstruction is inherently challenging due to the ill-posed nature of the inverse problem, which admits infinitely many solutions without additional constraints. In this study, we use the best available perturbative input, namely the perturbative spectral function at next-to-leading order for thermal SU(3)~\cite{Zhu:2012be,Vuorinen:2015wla}, to constrain the ultraviolet part of the spectral function.
The low-frequency infrared (IR) behavior is modeled using a Lorentzian-type ansatz, motivated by hard thermal loop (HTL) resummation techniques~\cite{Aarts:2002cc}. The transport coefficients are then determined from the slope of the transport peak at vanishing frequency via Eq.~(\ref{eq:kubo}). We emphasize that, beyond the intrinsic ill-posedness, the spectral reconstruction faces an additional challenge: the perturbative UV contribution to the spectral function scales as $\sim \omega^4$ and suppresses the sensitivity of the correlators to the transport peak. Consequently, precise correlators at large imaginary times $\tau T$, where IR physics dominates, are essential for constraining the transport behavior. We will return to this point in Sec.~\ref{sec:spectrum}.

\section{Computation of energy-momentum tensor correlators}
\label{sec:comp-emt}
\subsection{Lattice setup}
\label{sec:latt}

We compute the EMT correlators in four-dimensional SU(3) Yang-Mills theory on the lattice at seven temperatures: 0.76$\,T_c$, 0.9$\,T_c$, 1.125$\,T_c$, 1.27$\,T_c$, 1.5$\,T_c$, 1.9$\,T_c$, and 2.25$\,T_c$. At each temperature, configurations are generated using the standard Wilson gauge action~\cite{Wilson:1974sk} at three different lattice spacings to enable a controlled continuum extrapolation. The finest lattice spacing is $a=0.01164$ fm and the largest spatial extent is $L=aN_\sigma=3.31$ fm. The aspect ratio is not smaller than 4 across all ensembles to ensure that the volume effects are under control. The details of the lattice setup, including the temperature $T$,  lattice spacing $a$, spatial lattice extent $N_\sigma$, temporal extent $N_\tau$, $\beta$ value, and number of configurations, are summarized in Table~\ref{tab:lat_setup}. We generate $\sim$5000 configurations for each ensemble to achieve the desired statistical precision. The configurations are generated using a combination of the heat bath (HB)~\cite{Vladikas:1985uv} and over-relaxation (OR) algorithms~\cite{Creutz:1987xi,Adler:1987ce}. To reduce autocorrelations in Monte Carlo time, configurations are sampled every 500 sweeps, with each sweep consisting of one HB step and four OR steps. The first 4000 sweeps are discarded to ensure thermalization. In Table~\ref{tab:lat_setup}, we also include the parameter $n_\sigma$, which denotes the bin size used in the blocking method that complements gradient flow in improving the signal (see~\cite{Altenkort:2021jbk} for further discussion). 
The lattice spacing $a$ is set using the Sommer parameter $r_0$~\cite{Sommer:1993ce} with $r_0T_c = 0.7457$~\cite{Francis:2015lha}. The coefficients of the Allton-type ansatz used to set the scale were first determined in Ref.~\cite{Francis:2015lha} and later updated in Ref.~\cite{Burnier:2017bod}. The error analysis of this work is performed using bootstrap resampling with 1000 samples for each ensemble. The final results are quoted as the median and 68\% confidence intervals of the sample distribution. 

\begin{table}[htb]       
    \centering
    \begin{tabular}{cccccccc}                            
    \hline \hline
    $T/T_c$ & $a$ (fm) & $a^{-1}$ (GeV) & $N_{\sigma}$ & $n_{\sigma}$ & $N_{\tau}$ & $\beta$ & \#Conf. \tabularnewline
    \hline
    \multirow{3}{*}{0.76} & 0.03446 & 5.726 & 96 & 4 & 24  & 6.6506 & 6540
    \tabularnewline
    & 0.02757 & 7.157 & 120 & 6 & 30  & 6.8268 & 5693
    \tabularnewline
    & 0.02298 & 8.588 & 144 & 8 & 36  & 6.9742 & 5000
    \tabularnewline
    \hline
    \multirow{3}{*}{0.90} & 0.02910 & 6.780 & 96 & 4 & 24  & 6.7837 & 5000
    \tabularnewline
    & 0.02328 & 8.476 & 120 & 6 & 30  & 6.9634 & 5000
    \tabularnewline
    & 0.01940 & 10.171 & 144 & 8 & 36  & 7.1131 & 4998
    \tabularnewline
    \hline
    \multirow{3}{*}{1.125} & 0.02328 & 8.476 & 96 & 4 & 24  & 6.9634 & 5000
    \tabularnewline
    & 0.01862 & 10.594 & 120 & 6 & 30  & 7.1469 & 5000
    \tabularnewline
    & 0.01552 & 12.713 & 144 & 8 & 36  & 7.2989 & 5000
    \tabularnewline
    \hline
    \multirow{3}{*}{1.27}  & 0.02068 & 9.543 & 96 & 4 & 24  & 7.0606 & 5000
    \tabularnewline
    & 0.01654 & 11.93 & 120 & 6 & 30  & 7.2456 & 5000
    \tabularnewline
    & 0.01379 & 14.31 & 144 & 8 & 36  & 7.3986 & 5000
    \tabularnewline
    \hline
    \multirow{3}{*}{1.5}  & 0.01746 & 11.30 & 96 & 4 & 24  & 7.2005 & 5000
    \tabularnewline
     & 0.01397 & 14.13 & 120 & 6 & 30  & 7.3874 & 5000
    \tabularnewline
     & 0.01164 & 16.95 & 144 & 8 & 36  & 7.5416 & 5000
    \tabularnewline
    \hline
    \multirow{3}{*}{1.9}  & 0.02068 & 9.543 & 96 & 4 & 16  & 7.0606 & 5000
    \tabularnewline
    & 0.01654 & 11.93 & 120 & 6 & 20  & 7.2456 & 5000
    \tabularnewline
    & 0.01379 & 14.31 & 144 & 8 & 24  & 7.3986 & 5000
    \tabularnewline
    \hline
    \multirow{3}{*}{2.25}  & 0.01746 & 11.30 & 96 & 4 & 16  & 7.2005 & 5000
    \tabularnewline
    & 0.01397 & 14.13 & 120 & 6 & 20  & 7.3874 & 5000
    \tabularnewline
    & 0.01164 & 16.95 & 144 & 8 & 24  & 7.5416 & 5000
    \tabularnewline
    \hline \hline
    \end{tabular}
    \caption{Lattice setup used in the computation of EMT correlators, including the temperatures $T$, lattice spacing $a$,  lattice extents  $N_\sigma^3 \times N_\tau$, bin size $n_\sigma$ in the blocking method and the number of configurations.}
    \label{tab:lat_setup}
\end{table}

\subsection{Gradient flow}
The generated configurations are evolved according to the flow equations in Eq.~(\ref{eq:flow1}). A leading-order perturbative solution shows that the flowed gauge field $B_{\mu}(x,\tauf)$ and its unflowed counterpart 
$A_{\mu}(x)$ are related by a $\tauf$-dependent smearing~\cite{Luscher:2010iy},
\begin{align}
    \label{flow_solution}
        \begin{split}
        &B_{\mu}(x,\tauf)=\int \mathrm{d}^4y\ K_{\tauf}(x-y)A_{\mu}(y)\,,\\
        &K_{\tauf}(x)=\frac{e^{-x^2/4\tauf}}{(4\pi \tauf)^2}\,.
    \end{split}
\end{align}
This transformation is mediated by a Gaussian smearing kernel $K_{\tauf}$, which suppresses ultraviolet fluctuations over a characteristic length scale $\sqrt{8\tauf}$~\cite{Luscher:2011bx}. The smearing effectively regularizes short-distance fluctuations in the gauge fields, significantly enhancing the signal-to-noise ratio of the gauge field composite observables. 

On the lattice, gauge fields $A_{\mu}(x)$ are represented by gauge links $U(x,\mu)$. Accordingly, the flow equations are recast as
\begin{equation}
\begin{split}
V(x,\mu)|_{\tauf=0}&=U(x,\mu)\,, \\
    \frac{\mathrm{d}}{\mathrm{d}\tauf}V(x,\mu)&=-g_0^2 \{\partial_{x,\mu}S_G(V) \}V(x,\mu)\,,
    \end{split}
    \label{flowlat}
\end{equation}
where $V(x,\mu)$ denotes the flowed gauge link. The bare coupling $g_0$ is determined by $\beta=6/g_0^2$. $S_G$ is the gauge action of the flow, and in this work we adopt the Symanzik improved L$\ddot{\mathrm{u}}$scher-Weisz discretization~\cite{Luscher:1984xn, Luscher:1985zq}, which eliminates all the $\mathcal{O}(a^2)$ discretization effects originating from the discretized gradient flow equations or the gradient flow observables~\cite{Ramos:2015baa}.

We then evaluate the bare shear channel correlators $\langle UU\rangle$ and bulk channel correlators $\langle EE\rangle$ on the flowed configurations at different flow times. The required flowed field strength tensor $G_{\mu\nu}(x,\tauf)$ is discretized as
\begin{align}
    \label{eq:Fmunu}
    G_{\mu\nu}(n) = -\frac{i}{8}\Big{(}Q_{\mu\nu}(n)-Q_{\nu\mu}(n)\Big{)}\,,
\end{align}
where $Q_{\mu\nu}(n)$ is sum of four square plaquettes composed of gauge links, see, e.g.,~\cite{Gattringer:2010zz}. Note that all operators are expressed in lattice units. Since the EMT operator must vanish in vacuum, we subtract the corresponding zero-temperature expectation value from finite-temperature measurements. For viscosities, disconnected contributions must also be removed. 
Combining them requires evaluating the operators exclusively at finite temperature. In the bulk channel, we compute $\langle EE\rangle_T - \langle E\rangle_T^2$, while in the shear channel only $\langle UU\rangle_T$ is needed, since the stress density vanishes in thermal equilibrium~\cite{Taniguchi:2016ofw}.

\subsection{Renormalization}
\label{sec:sec_renormalization}
The renormalization constants $c_1$ and $c_2$ have been determined up to two-loop and three-loop in the $\overline{\mathrm{MS}}$ scheme~\cite{Harlander:2018zpi,Iritani:2018idk},
\begin{equation}
\begin{split}
c_1(\tauf) & =
\frac{1}{g_{\msbar}^2(\mu)} \sum_{n=0}^{2} k_1^{(n)}(L(\mu,\tauf)) \Big{[} \frac{g_{\msbar}^2(\mu)}{(4 \pi)^2 } \Big{]}^n ,\\
c_2(\tauf) & =
\frac{1}{g^2_{\msbar}(\mu)} \sum_{n=1}^{3} k_2^{(n)}(L(\mu,\tauf)) \Big{[} \frac{g^2_{\msbar}(\mu)}{(4 \pi)^2 } \Big{]}^n,
\end{split}
\label{eq:c1c2}
\end{equation}
where $L(\mu, \tauf) \equiv \log(2\mu^2 e^{\gamma_E}\tauf)$. The renormalization scale is set to the conventional choice $\mu=1/\sqrt{8\tauf}$. The coefficients $k_1^{(n)}$ and $k_2^{(n)}$ can be found in Ref.~\cite{Iritani:2018idk}. The coupling constant in the $\overline{\mathrm{MS}}$ scheme $g^2_{\msbar}$ can be obtained from the gradient flow coupling $g^2_{\mathrm{flow}}$ via a three-loop matching relation~\cite{Harlander:2016vzb}
\begin{equation}
g^2_{\mathrm{flow}} = g^2_{\msbar} K_E(g^2_{\msbar})\,,
\label{eq:cubic_eq}
\end{equation}
where $K_E$ is a perturbative correction factor~\cite{Harlander:2016vzb}. The flowed coupling constant takes the form~\cite{Fodor:2012td,Hasenfratz:2019hpg}
\begin{align}
g_{\mathrm{flow}}^2 = \frac{128\pi^2}{3(N_c^2-1)}\frac{1}{1+\delta(\tauf)}\langle \tauf^2 E\rangle\,,
\label{coupling_eq}
\end{align}
with $N_c = 3$, $E$ the flowed action density, and $\delta$ accounting for finite-volume corrections~\cite{Fodor:2012td,Hasenfratz:2019hpg}. The problem thus boils down to computing   $g_{\mathrm{flow}}^2$ in the vacuum. The lattice ensembles used for this calculation are listed in Table~\ref{tab:c2}. The flowed coupling constants are measured at different $\beta$ values that allow us to extrapolate $c_1$ and $c_2$ to the continuum limit. 

\begin{table}[htb]
        \centering
        \begin{tabular}{cccccc}
        \hline \hline
        $\beta$ & $a$[fm] ($a^{-1}$[GeV]) & $N_\tau$ & $N_{\sigma}$ & $T/T_c$ & \#Conf.\\ \hline
        7.3874 & 0.01397 (14.13) & 120 & 96 & 0.38 & 500 \\
        7.3986 & 0.01379 (14.31) & 120 & 96 & 0.38 & 500 \\
        7.5416 & 0.01164 (16.95) & 144 & 96 & 0.38 & 500 \\ \hline \hline
        \end{tabular}
        \caption{Lattice setup used to generate configurations at temperatures well below the confinement/deconfinement phase transition for the computation of the renormalization constants $c_1$ and $c_2$.}
        \label{tab:c2}
    \end{table}

\begin{figure*}[tbh]
\centerline{
\includegraphics[width=0.5\textwidth]{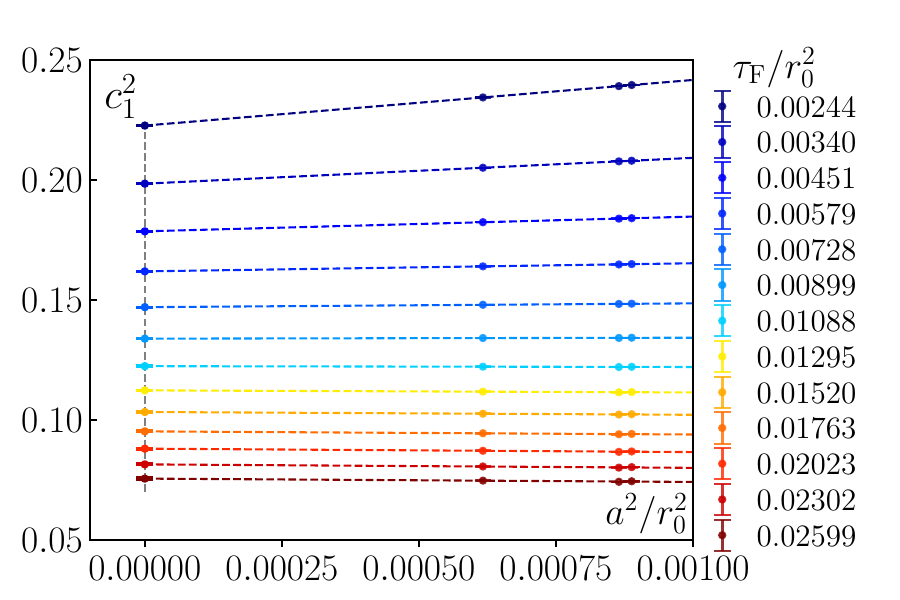}
\includegraphics[width=.5\textwidth]{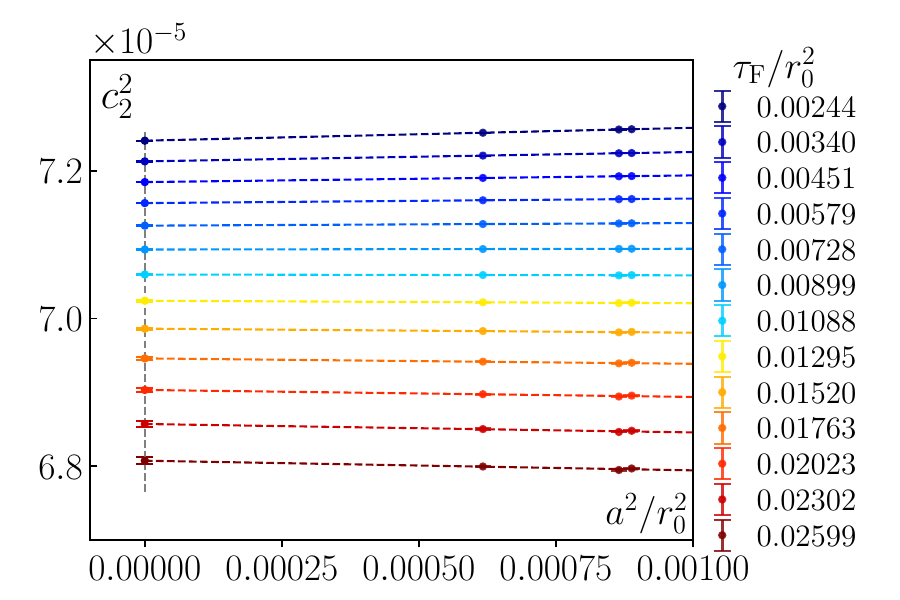}
}
\centerline{
\includegraphics[width=0.5\textwidth]{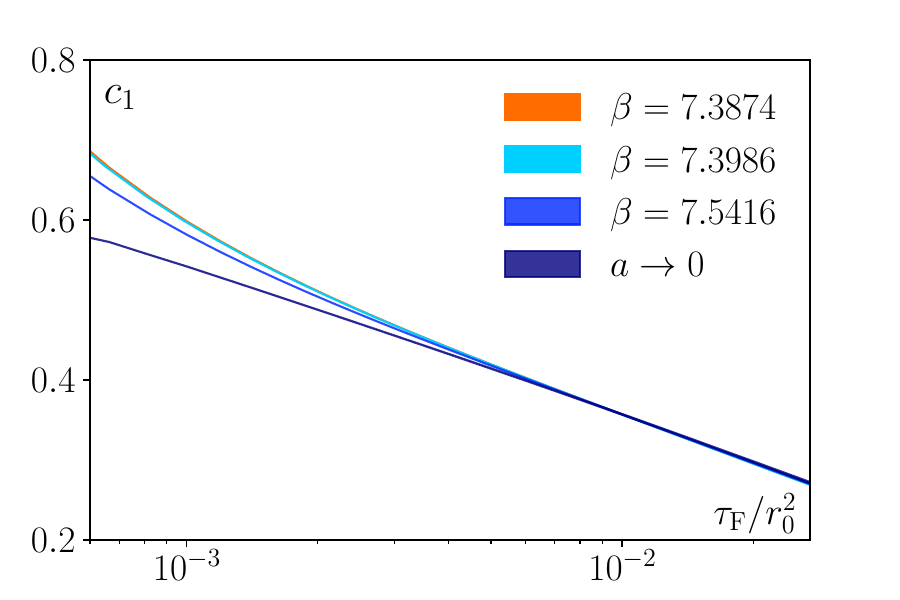}
\includegraphics[width=.5\textwidth]{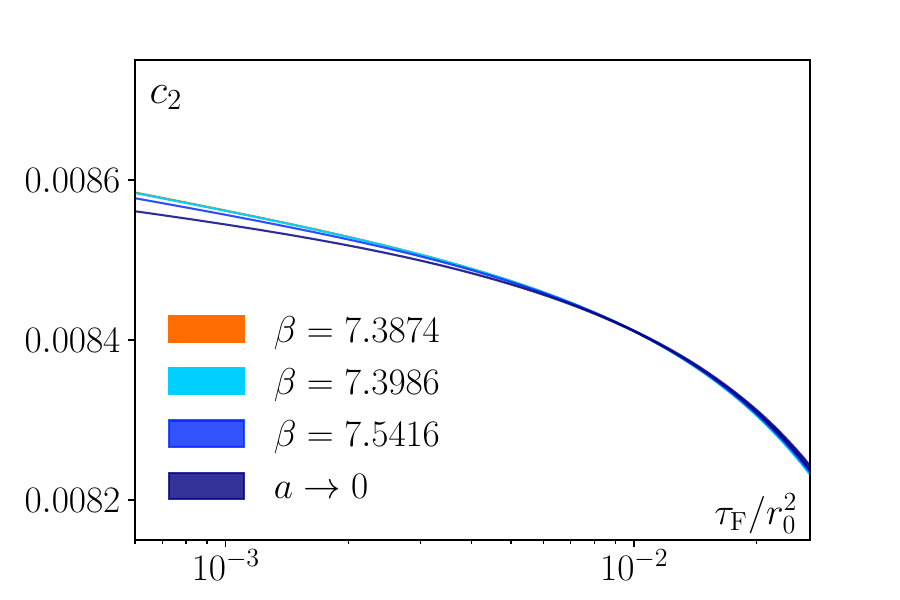}
}
\caption{Top: continuum extrapolation of $c_1$ (left) and $c_2$ (right) using Ans\"atze Eq.~(\ref{eq:c1_to_0}). Bottom: a comparison of $c_1$ (left) measured at finite lattice spacings and in the continuum limit and the same for $c_2$ (right).}
\label{fig:c1c2}
\end{figure*}

We note that the method of determining $c_1$ by comparing the flowed and unflowed entropy densities proposed in Ref.~\cite{Altenkort:2022yhb} is not applicable here. This is because our temperature range extends down to 0.76$\,T_c$. A precise determination of $c_1$ would require simulations spanning from high temperatures (e.g., $\sim 5T_c$ ) down to the target temperature (0.76$\,T_c$). However, lattice computations of the entropy density at low temperatures—particularly below $T_c$—are notoriously challenging, rendering $c_1$ obtained via this method imprecise. Instead, we determine both $c_1$ and $c_2$ from the coupling constant, which is better suited for our analysis.
 
The renormalization constants obtained at three  lattice spacings are extrapolated to the continuum limit using the following ansatz
\begin{equation}
c_{1,2}^2(a) = m~ (a/r_0)^2 + c_{1,2}^2(a=0)\,,
\label{eq:c1_to_0}
\end{equation}
where $m$ and $c_{1,2}^2(a=0)$ are fit parameters. The top panels of Fig.~\ref{fig:c1c2} show the extrapolations for $c_1^2$ (left) and $c_2^2$ (right). It can be seen that the ansatz describes the data well. A comparison of the renormalization constants at finite lattice spacing and in the continuum limit is shown in the bottom panels of Fig.~\ref{fig:c1c2}. The overlap of uncertainty bands at nonsmall $\tauf/r_0^2$ (note that the $x$-axis is in log scale) indicates that lattice spacing effects are negligible. We have also performed extrapolations assuming $\mathcal{O}(a^4)$ discretization error and find that, relative to the $\mathcal{O}(a^2)$-error ansatz in Eq.~(\ref{eq:c1_to_0}), the difference is less than 1\% for both $c_1$ and $c_2$ at most of the discrete flow times within $\tauf/r_0^2\in [0, 0.0265]$, a window we will address in the next section.

\subsection{Analysis of the EMT correlators}
\label{sec:extrapolate}

The bare EMT correlators $\langle UU\rangle$ and subtracted $\langle EE\rangle$, with $c_1$ and $c_2$ scaled out, are computed under gradient flow, incorporating the blocking method to further improve the signal-to-noise ratio. In the blocking method, for each temporal distance $\tau$, correlators are sampled at all possible spatial distances. At short spatial distances the signal dominates, while the noise is distributed uniformly across all distances. The blocking method leverages this by replacing spatial correlators at distances where the signal is weak by values obtained from fits, performed within a suitable spatial window. These fits employ an appropriate ansatz to model the decay of the signal. In this work, the fit model is given by shear and bulk channel correlators computed at leading order in perturbative theory in discretized space~\cite{Altenkort:2022yhb}, rescaled by an overall factor as a fit parameter. Further technical details about the blocking method and fits can be found in the pioneering work~\cite{Altenkort:2021jbk,Altenkort:2022yhb}. In contrast to the original proposal of Refs.\cite{Altenkort:2021jbk,Altenkort:2022yhb}, which neglects correlations among different spatial distances, we explicitly account for these correlations by performing correlated blocking fits using the covariance matrix constructed from data at different spatial separations. In case the covariance matrix becomes ill conditioned, we regularize it by adding a scaled identity matrix~\cite{golub1999tikhonov},
\begin{align}
C + \alpha  C^{\mathrm{max}}_{ii}\mathbf{1} \longrightarrow C\,,
\end{align}
where the scaling factor $\alpha$ is  increased incrementally from 0 until the resulting $p$-value exceeds 0.05, and $C^{\mathrm{max}}_{ii}$ denotes the largest diagonal element. 

The bare correlators evaluated on the lattice suffer from discretization effects. To mitigate these effects, tree-level improvement is applied by multiplying the nonperturbative EMT correlator measured on the lattice, $G_{\rm lat}(\tau T)$, by the ratio of its LO perturbative counterpart in the continuum, $G^{\mathrm{LO}}_{\mathrm{cont}}(\tau T)$, to the same LO perturbative correlator calculated on the lattice, $G^{\mathrm{LO}}_{\mathrm{lat}}(\tau T)$~\cite{Gimenez:2004me,Meyer:2009vj},
\begin{equation}
\label{ratio_tt}
G^{\mathrm{t.l.}}(\tau T)=G_{\rm lat}(\tau T)   
~ \frac{G^{\mathrm{LO}}_{\mathrm{cont}}(\tau T) }{G^{\mathrm{LO}}_{\mathrm{lat}}(\tau T) }\, .
\end{equation}
The LO perturbative EMT correlators $G^{\mathrm{LO}}_{\mathrm{cont}}(\tau T)$ in the shear and bulk channels read~\cite{Meyer:2007ic,Meyer:2007dy}
\begin{equation}
\label{pert_cont}
\begin{split}
& \frac{G^{\mathrm{LO}}_{\mathrm{cont,shear}}(\tau T)}{T^5} = \frac{32d_A}{5\pi^2} \Big(f(x)-\frac{\pi^4}{72} \Big)\, , \\
& \frac{G^{\mathrm{LO}}_{\mathrm{cont,bulk}}(\tau T)}{T^5} = \frac{484d_A}{16\pi^6}g^4 \Big(f(x)-\frac{\pi^4}{60} \Big)\, ,
\end{split}
\end{equation}
where $x = 1 - 2 \tau T$, $f(x) = \int_0^\infty dr~ r^4  \cosh^2(x r)/\sinh^2 r$ and $d_A = 8$ is the number of gluons. The lattice LO perturbative correlators using clover discretization, $G^{\mathrm{LO}}_{\mathrm{lat}}(\tau T)$, can be evaluated numerically following the method outlined in Ref.~\cite{Meyer:2009vj}. The tree level-improved correlators are denoted as $G^{\mathrm{t.l.}}(\tau T)$. For clarity, we normalize these correlators by a  factor $G_{\rm{norm}}$, defined as $G^{\mathrm{LO}}_{\mathrm{cont,shear}}$ for the shear channel and $G^{\mathrm{LO}}_{\mathrm{cont,bulk}}/g^4$ for the bulk channel. In the latter case, the coupling constant factor $g^4$ is scaled out for convenience.

\begin{figure}[tbh]
    \centering
    \includegraphics[width=0.5\textwidth]{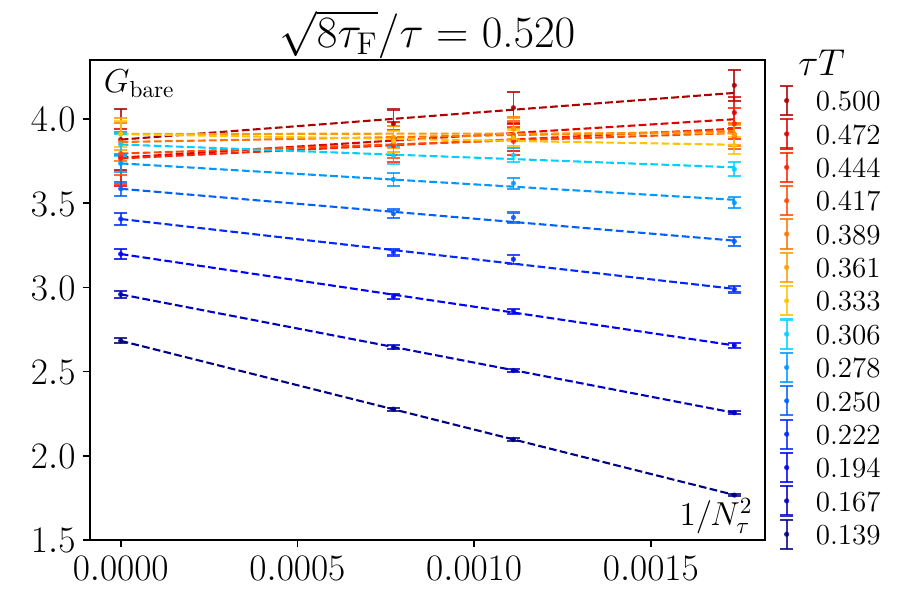}
    \caption{Continuum extrapolation of the bare EMT correlator in the shear channel at $\sqrt{8 \tauf} / \tau = 0.520$ and $T=0.9 \,T_c$ via a joint fit with the ansatz in Eq.~(\ref{eq:joints}).}
    \label{fig:a0_0.9}
\end{figure}

The normalized, tree level-improved correlators, denoted as $G_{\mathrm{bare}}(\tau T)$, must be extrapolated to the continuum limit to eliminate discretization effects and an underlying divergence of the form $a^2/\tauf$~\cite{Altenkort:2020fgs}. Since the dominant discretization error in this study originates from the Wilson gauge action and starts at  $\mathcal{O}(a^2)$, the continuum extrapolation ansatz must include at least a term linear in $1/N^2_{\tau}$. For this reason, we perform uncorrelated joint fits over all available $\tau T$ using the following $\tau T$-dependent data-driven model:
\begin{equation}
\begin{split}
G_{\mathrm{bare}}(N_\tau, \tau T)
   & = G_{\mathrm{bare}}(a=0, \tau T) \\
&+ \Big{(}m_0 + m_1~\tau T + \frac{m_2}{\tau T} \Big{)}/N_{\tau}^{2}\, .
\end{split}
\label{eq:joints}
\end{equation}
This model generalizes the simpler form that retains only the first two terms commonly used in fits performed separately for each $\tau T$. The general model in Eq.~(\ref{eq:joints}) allows the slope to vary with $\tau T$ when used in a joint fit. In Appendix~\ref{app:cont-joint-sepa} we provide the details of the joint fits, including a comparison between joint fits using the model in Eq.~(\ref{eq:joints}) and separate fits using the simpler model. We find that the results are consistent within 1-$\sigma$ statistical uncertainties for all $\tau T$ within the ``valid" flow-time window discussed below. Figure~\ref{fig:a0_0.9} illustrates the continuum extrapolation for the bare EMT correlator in the shear channel at $\sqrt{8 \tauf} / \tau = 0.520$ and $T=0.9\,T_c$. 
 
\begin{figure*}[tbh] 
\centerline{
\includegraphics[width=0.5\linewidth]{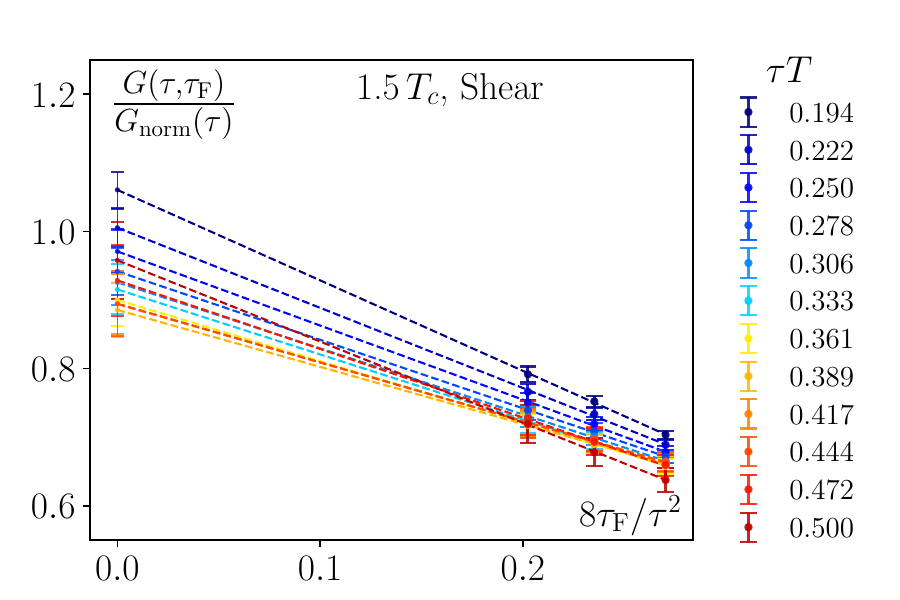}
\includegraphics[width=0.5\linewidth]{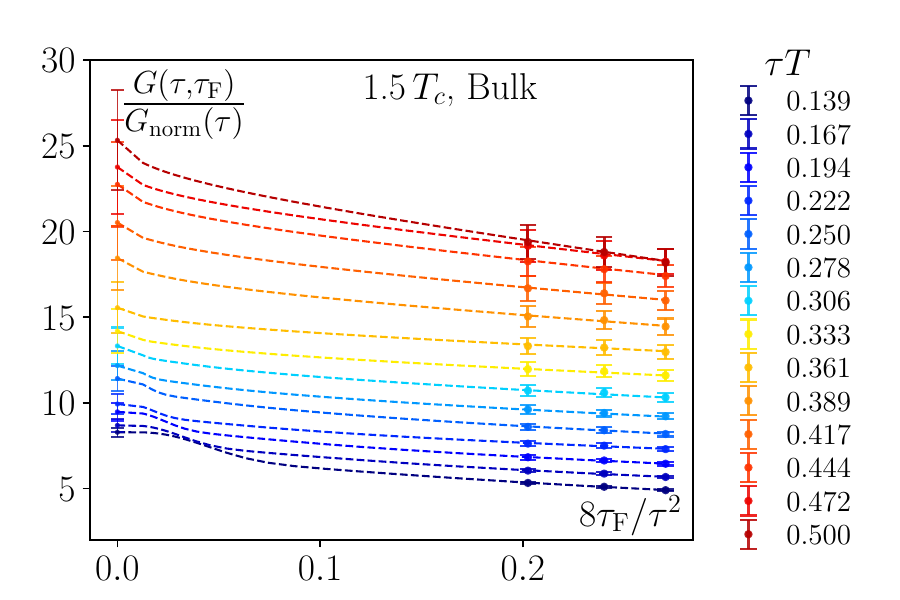}
}
\caption{Left: flow-time extrapolation for the shear channel correlators at $T=1.5\,T_c$. Right: same as the left panel but for the bulk channel.}
\label{fig:flow-extrap}
\end{figure*}
The bare continuum-extrapolated correlators $G_{\mathrm{bare}}(\tau T)$ are multiplied by the corresponding renormalization constants $c_1$ or $c_2$, which are also extrapolated to the continuum limit in Sec.~\ref{sec:sec_renormalization}, to construct the complete correlators. The complete correlators are then extrapolated to the zero flow-time limit to recover the correctly renormalized operators. There are two main considerations in the $\tauf\rightarrow 0$ extrapolation. First, one must identify a proper flow-time window:
 the flow time must be large enough to suppress UV fluctuations, yet small enough that the flow smearing remains controlled. To this end, Ref.~\cite{Altenkort:2020axj} proposed the window $\sqrt{8\tauf}/\tau \in [0.37, 0.52]$, which has proven effective. We adopt the upper bound and increase the lower bound to 0.45 to better suppress the slight nonlinearity around $\sqrt{8\tauf}/\tau \sim 0.43$ observed in our data. Second, since the field strength tensors used to construct the EMT correlators are discretized with clovers, the target operator has an effective radius of $\sqrt{2}a$. To ensure that the flow radius $\sqrt{8\tauf}$ extends beyond the operator size, we require $\sqrt{8\tauf} \geq \sqrt{2}a$.

Like the bare correlators, the renormalization constants are also subject to lattice spacing effects. Controlling these effects therefore places additional constrains on the admissible range of flow times. To this end, we compute the ratio of the renormalization constants obtained on coarser lattices ($\beta=7.3874$ and 7.3986) to those on the finest lattice ($\beta=7.5416$). These discretization effects can be quantified through deviations of the ratio from unity. It turns out that, for $c_1$, these effects are strongly suppressed at large flow time but remain significant at very small flow time. To ensure that these effects do not exceed the statistical uncertainties of the corresponding bare correlators, we require the deviations to be smaller than the maximum statistical errors of the bare correlators at $\sqrt{8\tauf}/\tau=0.45$ and $\tau T=0.5$. We choose $\sqrt{8\tauf}/\tau=0.45$ for conservative error estimation, as errors are larger at smaller flow times. The point $\tau T=0.5$ is prioritized due to its relevance to transport physics and typically it also carries the largest uncertainty. We find that the $\beta=6.6506$ lattice at $0.76\,T_c$ yields the largest statistical error (3.791\%) for $\langle UU\rangle$, imposing an additional constraint on the flow time in the shear channel: $\tauf/r_0^2 \geq 0.0002$.

In contrast, lattice spacing effects for $c_2$ are very small: the ratios of $c^2_2$ between different lattice spacings deviate by less than 0.8\% across all $\tauf/r_0^2$ values.  This is evident from the significantly smaller $y$-axis scale in the bottom right panel of Fig.~\ref{fig:c1c2} compared to the bottom left panel. Since 0.8\% is much smaller than the statistical errors of the bare $\langle EE\rangle$ correlators, we neglect the flow-time constraint stemming from $c_2$.

\begin{figure*}[tbh] 
 \centerline{
 \includegraphics[width=0.5\textwidth]{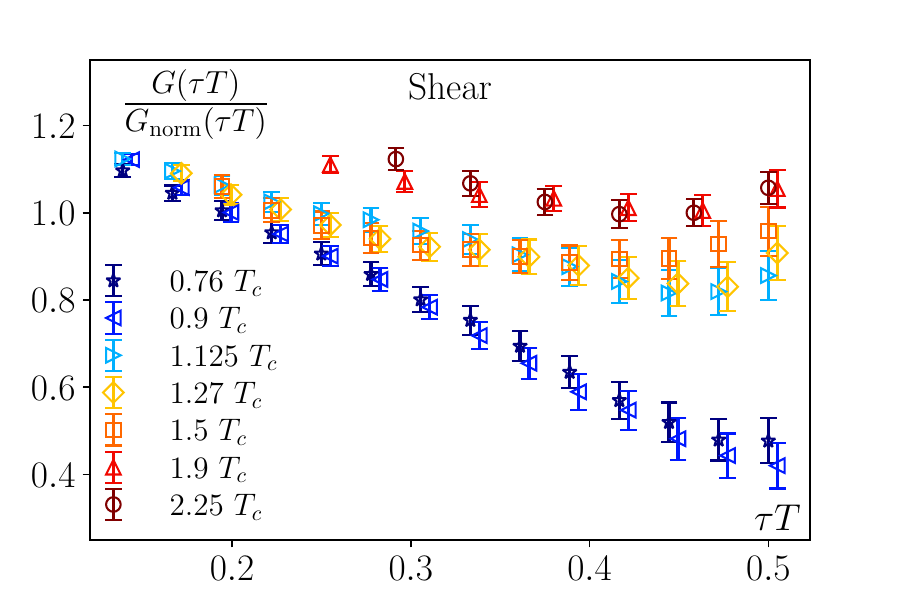}
 \includegraphics[width=0.5\textwidth]{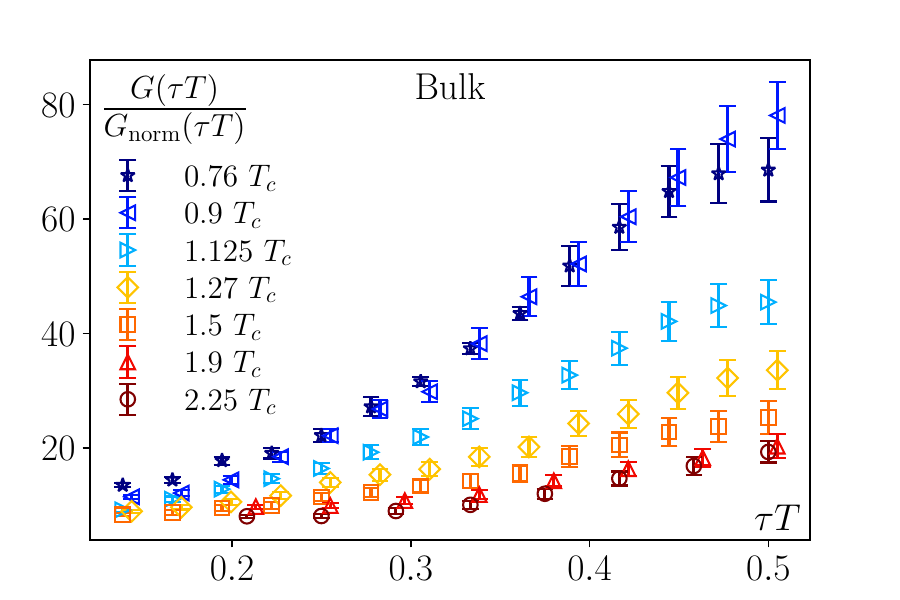}
 }
 \caption{Left: shear channel correlators after the double extrapolation for all temperatures. Right: same as the left panel but for the bulk channel.}
 \label{fig:final-corr}
\end{figure*}

We next perform the $\tauf \rightarrow 0$ extrapolation. Inspired by small flow-time expansions of operators built from flowed gauge fields~\cite{Luscher:2011bx}, shear correlators exhibit corrections that are polynomial in $\tauf$  as $\tauf \rightarrow 0$. Accordingly, for each $\tau T$ we perform the $\tauf\rightarrow0$ extrapolation  using 
\begin{equation}
\begin{split}
G(\tauf/\tau^2) = G(\tauf=0)+m' ~ \frac{\tauf}{\tau^2}\, .
\end{split}
\label{eq:flow-extrap-shear}
\end{equation}
For the bulk channel, we use a different extrapolation form. This is because the bulk correlators are essentially the correlation of two flowed trace anomaly operators $\theta$, whose flow-time dependence in three-loop perturbation theory is given by~\cite{Suzuki:2021tlr}
\begin{align}
    \theta(\tauf)=\Big{(}1-c \big{(}\frac{g^2(\mu(\tauf))}{(4 \pi)}\big{)}^{3}\Big{)} \theta(\tauf=0)\, . 
    \label{eq:flow-extrap-bulk}
\end{align}
Therefore, we take the square root of the bulk correlators and fit it to Eq.~(\ref{eq:flow-extrap-bulk}) separately for each $\tau T$. The flow-time extrapolation is illustrated in Fig.~\ref{fig:flow-extrap} for both channels, taking $T=1.5\,T_c$ as an example. For the shear and bulk channels, the fits yield $\chi^2$/degree of freedom(d.o.f.) in the ranges 0.003--0.304 and 0.002--1.127, respectively, across all $\tau T$ and temperatures. The double-extrapolated correlators at different temperatures are summarized in Fig.~\ref{fig:final-corr}. Both channels show clear temperature dependence, particularly at large $\tau T$, indicating temperature-dependent changes in the viscosities that we investigate in the next section.

\section{Spectral analysis}
\label{sec:spectrum}
The double-extrapolated correlators obtained in the previous section are used to reconstruct the corresponding spectral function by inverting the convolution equation in Eq.~(\ref{eq:correlator-spf}). Since the lattice correlators are available only at a finite set of discrete temporal separations $\tau$, while the target spectral function is continuous, there are an infinite number of possible solutions. To address this ill-posed problem, numerous methods have been proposed and widely adopted, including the maximum entropy method~\cite{asakawa2001maximum}, the Backus-Gilbert method~\cite{Backus1968}, the stochastic optimization method~\cite{Ding:2017std}, the Tikhonov regularization~\cite{dudal2014kallen}, the Bayesian approach~\cite{Burnier:2013nla}, the sparse modeling approach~\cite{Itou:2020azb}, and others~\cite{Chen:2021giw}. Beyond these systematic frameworks, a simple $\chi^2$ minimization proves effective and robust when guided by a model with a solid theoretical basis~\cite{Altenkort:2022yhb}. In this work, we adopt this strategy and model the UV regime of the spectral function using the corresponding NLO spectral function computed for the SU(3) thermal medium~\cite{Zhu:2012be,Laine:2011xm}
\begin{align}
\label{rho_shear}               
    \rho_{\rm{shear}}^{\mathrm{pert}}(\omega) =& \frac{d_A \  \omega^4}{10\pi}
    \coth\Big{(}\frac{\omega}{4T}\Big{)} - \frac{8}{9} d_A \omega^4 \coth\Big{(}\frac{\omega}{4T}\Big{)}  \nonumber \\
    &\times \frac{g^2(\bar{\mu})N_c}{(4\pi)^3}\, ,  \\
    \rho_{\rm{bulk}}^{\mathrm{pert}}(\omega) =& \frac{d_A c^2_{\theta}\  \omega^4g^4}{4\pi} \coth\Big{(}\frac{\omega}{4T}\Big{)} + d_A c^2_{\theta}\ \omega^4 \coth\Big{(}\frac{\omega}{4T}\Big{)}  \nonumber \\
    &\times \frac{g^6(\bar{\mu})N_c}{(4\pi)^3}\biggl[\frac{22}{3} \ln\frac{\bar{\mu}^2}{\omega^2} + \frac{73}{3} + 8\, \phi^{ }_T(\omega) \biggr]\, , \nonumber
\end{align}
where $d_A = 8$, and the dimensionless function $\phi_T(\omega)$ encodes the temperature dependence in the bulk channel~\cite{Zhu:2012be}. We omit the corresponding thermal term in the shear channel, $\phi_T^{\eta}(\omega)$, calculated in Ref.~\cite{Zhu:2012be}, for two reasons: (i) its contribution to the Euclidean correlator is negligible, and (ii) in the deep IR region---where its evaluation becomes less reliable---it exhibits an artificial divergence that strongly distorts the transport peak.  The coefficient $c_{\theta}$ can be found in Ref.~\cite{Zhu:2012be}.

\begin{figure}[tbh]
    \centering
    \includegraphics[width=0.5\textwidth]{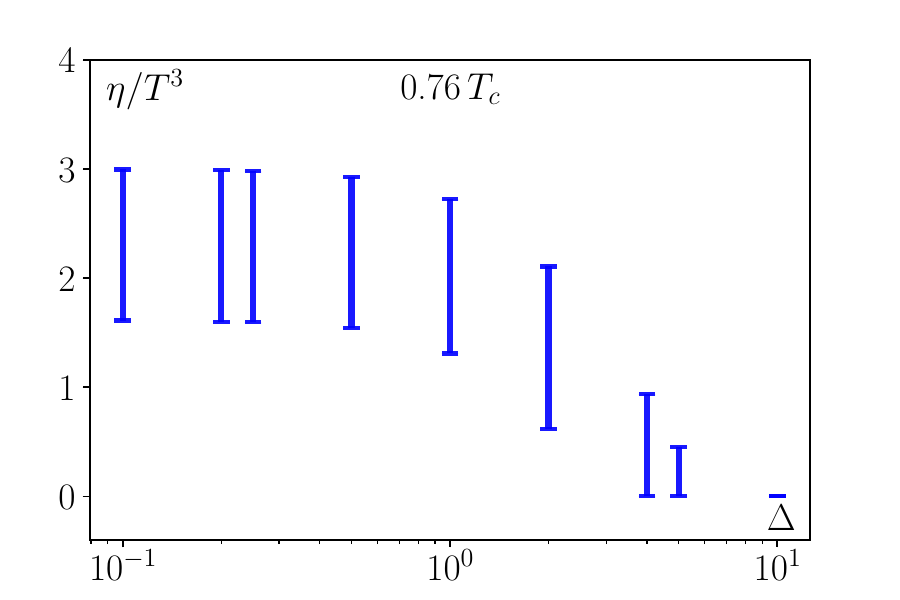}
    \caption{The dependence of $\eta/T^3$ on $\Delta$ at $0.76\,T_c$ with $C=1$.}
    \label{fig:0.76_eta_Delta}
\end{figure}

The running coupling must be evaluated at an appropriate scale. Previous detailed investigations~\cite{Zhu:2012be,Laine:2011xm} revealed that both channels exhibit a switching point for the running scale, denoted by $\mu_{\rm{swi}}$: below this point the scale is held fixed, while above it the scale runs with $\omega$~\cite{Kajantie:1997tt}. 
For the shear channel we use
\begin{equation}
\ln\left( \bar{\mu}_{\rm{shear}} \right)=
\begin{cases} \ln\left( 4\pi T\right) -\gamma_{\mathrm{E}} -\frac{1}{22},
& \text{if}\ \mu<\mu^{\rm{swi}}_{\rm{shear}} , \\ 
 \omega, &\text{if}\ \mu\geq\mu^{\rm{swi}}_{\rm{shear}} ,
\end{cases}
\label{eq:mu_shear}
\end{equation}
while for the bulk channel we use~\cite{Laine:2011xm}
\begin{equation}
\ln\left( \bar{\mu}_{\rm{bulk}} \right)=\begin{cases}\ \ln\left( 4\pi T\right) -\gamma_{\mathrm{E}} -\frac{1}{22},& \text{if}\ \mu<\mu^{\rm{swi}}_{\rm{bulk}}, \\ 
\ \ln\left(  \omega \right)  -\frac{73}{44}, &\text{if}\ \mu\geq\mu^{\rm{swi}}_{\rm{bulk}} .\end{cases}
\label{eq:mu_bulk}
\end{equation}
With these scale choices, we evaluate the coupling at one-loop order using $T_c=1.24\,\Lambda_{\overline{\rm{MS}}}$~\cite{Francis:2015lha} and the RunDec package~\cite{Herren:2017osy,Chetyrkin:2000yt}. Equating the first line and the second line of Eq.~(\ref{eq:mu_shear}) and Eq.~(\ref{eq:mu_bulk}), respectively, yields $\mu^{\rm{swi}}_{\rm{shear}}=2.146\pi T$ and $\mu^{\rm{swi}}_{\rm{bulk}}=11.28\pi T$.

\begin{figure*}[htb]
\centerline{
\includegraphics[width=0.33\textwidth]{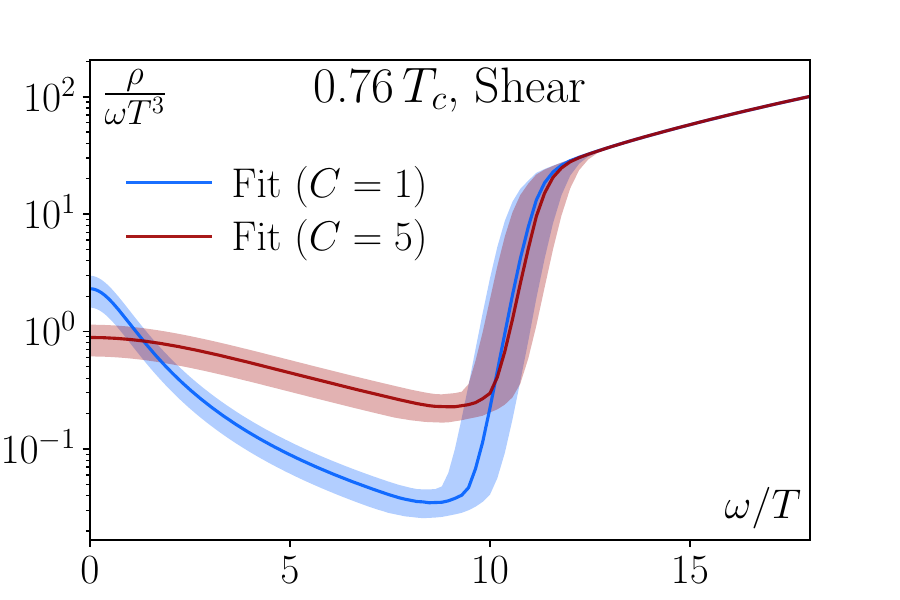}
\includegraphics[width=0.33\textwidth]{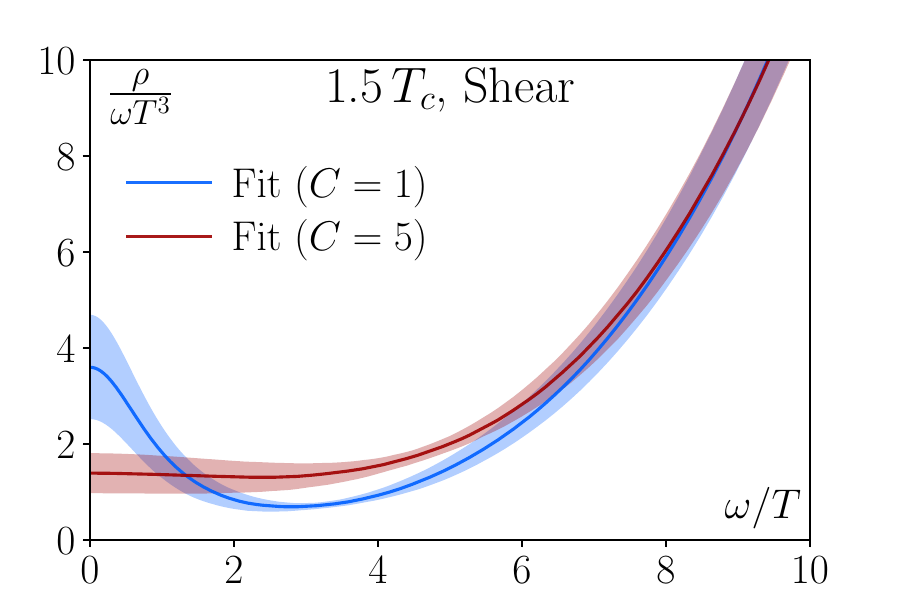}
\includegraphics[width=0.33\textwidth]{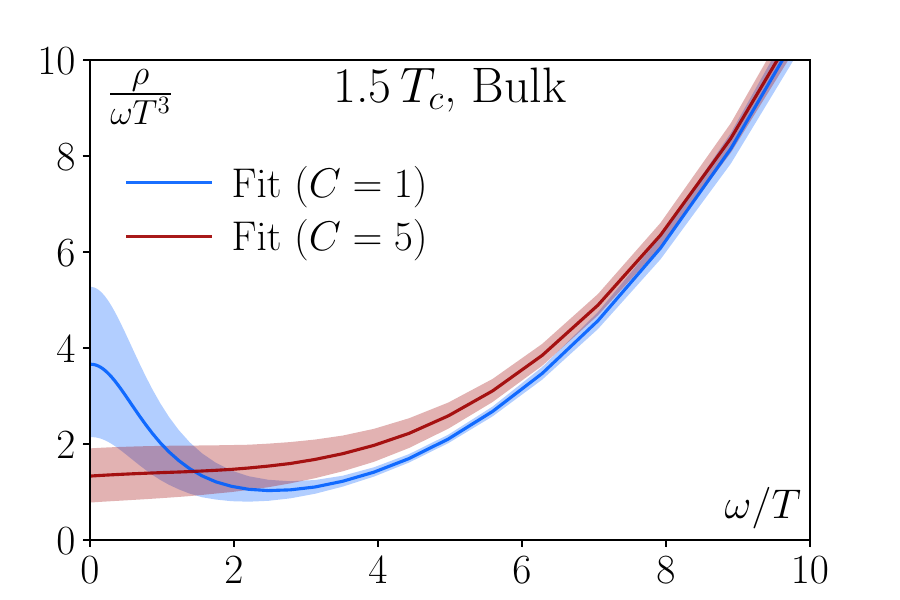}}
\centerline{
\includegraphics[width=0.33\textwidth]{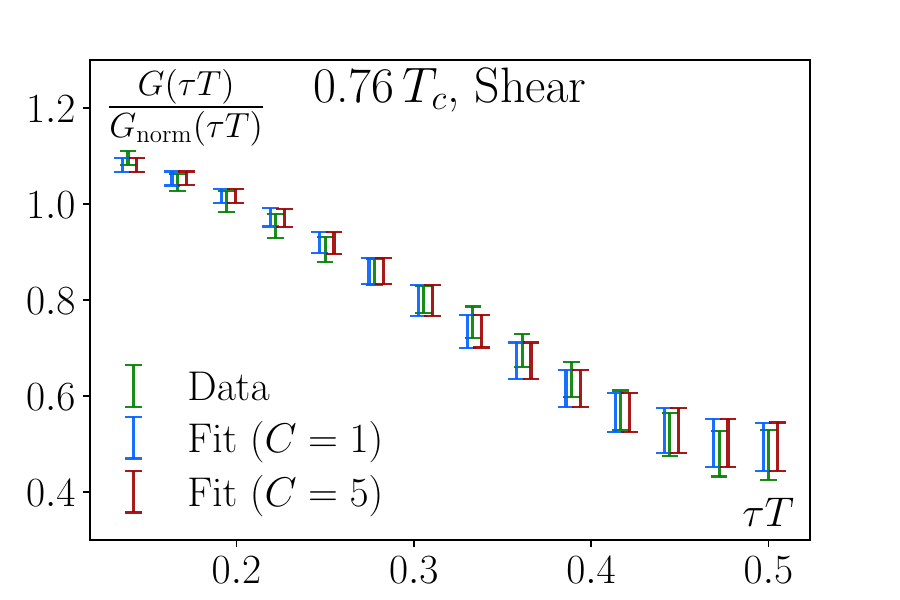}
\includegraphics[width=0.33\textwidth]{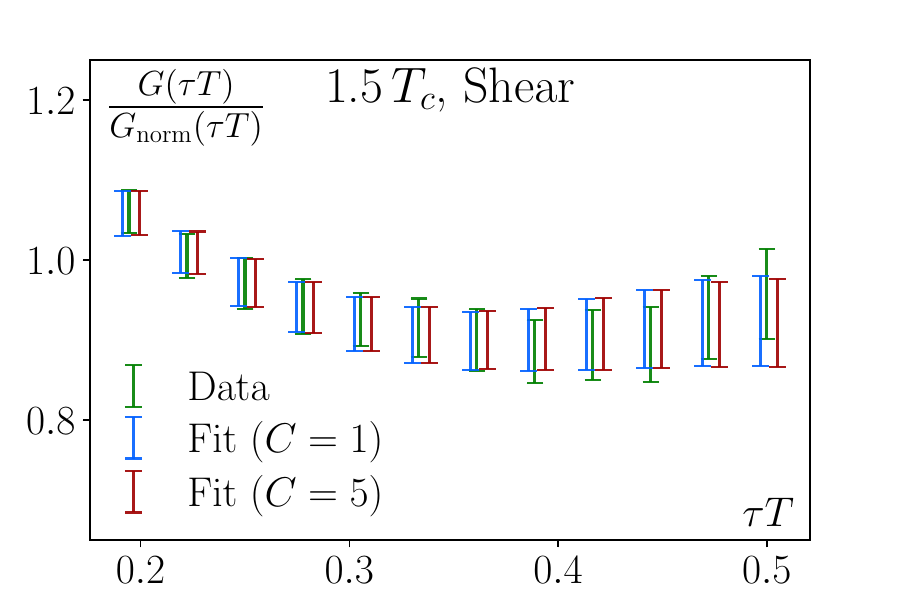}
\includegraphics[width=0.33\textwidth]{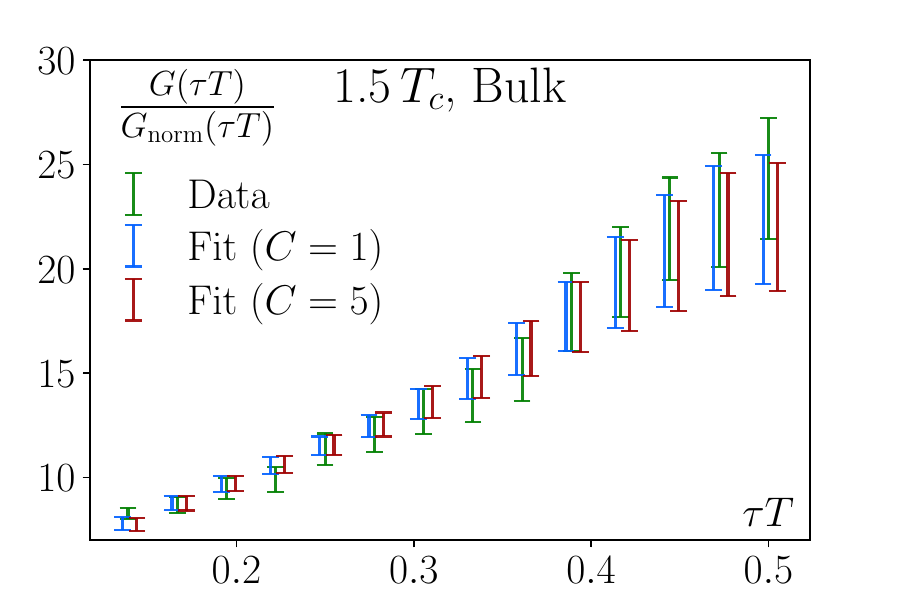}
}
\caption{Top left: fitted spectral functions at 0.76$\,T_c$ in the shear channel. Top middle: same as the top left panel but for 1.5$\,T_c$. Top right: same as the top middle panel but for the bulk channel. The bands denote the statistical uncertainties. Bottom: the fitted correlators corresponding to the top panels. }
\label{fig:spf_shear_bulk}
\end{figure*}

The form of the IR part of the spectral function is not known \textit{a priori}. However, it is widely accepted that this region can be modeled using a Lorentzian transport peak. Combining the IR and UV Ans\"atze, we arrive at a complete model for the spectral function,
\begin{equation}
\frac{\rho_{\mathrm{model}}(\omega)}{\omega T^3}=\frac{A}{T^3}\frac{C^2}{C^2+(\omega/T)^2}+B\frac{\rho^{\mathrm{pert}}(\omega)}{\omega T^3}\, ,
\label{eq:spf_model}
\end{equation}
where $A$ represents the transport contribution proportional to viscosities, and $B$ accounts for uncertainties in the renormalization of the perturbative spectral function. Both $A$ and $B$ are fit parameters. The width of the Lorentzian peak, $C$, characterizes the lifetime of thermal excitations in a strongly coupled medium and cannot be fixed reliably without additional input. We therefore bracket this dominant modeling uncertainty by fixing $C$ to either $1$ or $5$, corresponding to an extremely sharp or broad transport peak, respectively.

We note that for the shear channel, an anomalous dimension is needed to improve the fit quality by replacing $\omega^4$ with $\omega^{4+\gamma}$ in the first line of Eq.~(\ref{rho_shear}). For the bulk channel, a thermal sum rule requires subtracting half a constant-in-$\tau$ contribution from the correlators due to the presence of a $\delta$ peak in the spectral function, $\rho /(\omega T^3) = \pi \frac{\epsilon + p}{T^4} \frac{(3c_s^2 - 1)^2}{c_s^2}\delta(\frac{\omega}{T})$, which disappears for $T<T_c$~\cite{Meyer:2011gj}. The entropy density $(\epsilon+p)/T^4$ and speed of sound $c_s^2$ are taken from Ref.~\cite{Giusti:2025fxu}. We also note that three models were adopted in the previous work~\cite{Altenkort:2022yhb}, whereas here we use only the last model with two distinct width parameters. This is because the first two models of Ref.~\cite{Altenkort:2022yhb} effectively reproduce the present setup after varying the transport peak width.

\begin{table*}[t]
\centering
\renewcommand{\arraystretch}{1.5}
\begin{tabular*}{\textwidth}{@{\extracolsep{\fill}}cccccccccc}
\hline\hline
\multicolumn{2}{c}{\multirow{2}{*}{}} & \multicolumn{5}{c}{Shear Channel} & \multicolumn{3}{c}{Bulk Channel} \\
\cline{3-10}
$T/T_c$ & $C$ & $A/T^3$ & $B$ & $\gamma$ & $\omega_0/T$ & $\chi^2/\text{d.o.f.}$ & $A/T^3$ & $B$ & $\chi^2/\text{d.o.f.}$ \\
\hline
\multirow{2}{*}{0.76} & 1 & \err{2.312}{0.669}{0.717} & \err{0.076}{0.001}{0.001} &  & \err{11.094}{0.674}{0.702} & 0.355 & \err{22.956}{1.958}{2.121} & \err{0.606}{0.016}{0.015} & 1.053 \\
& 5 & \err{0.889}{0.253}{0.276} & \err{0.076}{0.001}{0.001} &  & \err{11.253}{0.697}{0.724} & 0.343 & \err{8.104}{0.685}{0.749} & \err{0.590}{0.017}{0.015} & 0.996 \\
\hline
\multirow{2}{*}{0.9} & 1 & \err{2.422}{0.719}{0.668} & \err{0.078}{0.001}{0.001} &  & \err{11.975}{0.585}{0.605} & 0.634 & \err{28.645}{3.034}{3.726} & \err{0.639}{0.024}{0.022} & 1.315 \\
& 5 & \err{0.923}{0.270}{0.256} & \err{0.078}{0.001}{0.001} &  & \err{12.121}{0.602}{0.620} & 0.611 & \err{10.283}{1.115}{1.330} & \err{0.616}{0.023}{0.024} & 1.014 \\
\hline
\multirow{2}{*}{1.125} & 1 & \err{0.745}{0.938}{0.745} & \err{0.044}{0.007}{0.006} & \err{0.156}{0.045}{0.044} &  & 0.167 & \err{9.917}{2.218}{2.267} & \err{0.700}{0.023}{0.024} & 0.622 \\
& 5 & \err{0.278}{0.361}{0.278} & \err{0.044}{0.007}{0.007} & \err{0.160}{0.046}{0.046} &  & 0.164 & \err{3.597}{0.800}{0.825} & \err{0.689}{0.024}{0.024} & 0.424 \\
\hline
\multirow{2}{*}{1.27} & 1 & \err{2.196}{0.918}{1.109} & \err{0.032}{0.009}{0.007} & \err{0.252}{0.074}{0.076} &  & 0.191 & \err{6.191}{1.852}{1.932} & \err{0.711}{0.024}{0.024} & 0.513 \\
& 5 & \err{0.841}{0.352}{0.428} & \err{0.031}{0.009}{0.007} & \err{0.265}{0.078}{0.079} &  & 0.188 & \err{2.204}{0.676}{0.688} & \err{0.704}{0.024}{0.024} & 0.541 \\
\hline
\multirow{2}{*}{1.5} & 1 & \err{3.599}{1.093}{1.080} & \err{0.023}{0.010}{0.007} & \err{0.350}{0.114}{0.116} &  & 0.122 & \err{3.594}{1.614}{1.515} & \err{0.753}{0.030}{0.029} & 0.694 \\
& 5 & \err{1.393}{0.415}{0.418} & \err{0.021}{0.010}{0.007} & \err{0.379}{0.120}{0.125} &  & 0.126 & \err{1.268}{0.579}{0.551} & \err{0.750}{0.031}{0.030} & 0.778 \\ 
\hline
\multirow{2}{*}{1.9} & 1 & \err{3.248}{1.036}{1.156} & \err{0.033}{0.018}{0.012} & \err{0.278}{0.159}{0.147} &  & 0.248 & \err{2.269}{0.984}{1.120} & \err{0.848}{0.036}{0.035} & 1.170 \\
& 5 & \err{1.286}{0.401}{0.452} & \err{0.030}{0.018}{0.012} & \err{0.311}{0.165}{0.161} &  & 0.253 & \err{0.817}{0.361}{0.413} & \err{0.842}{0.038}{0.036} & 1.271 \\
\hline
\multirow{2}{*}{2.25} & 1 & \err{4.946}{1.031}{1.127} & \err{0.010}{0.009}{0.005} & \err{0.704}{0.240}{0.217} &  & 0.564 & \err{2.860}{0.952}{0.928} & \err{0.795}{0.034}{0.034} & 1.710 \\
& 5 & \err{1.946}{0.386}{0.433} & \err{0.008}{0.008}{0.004} & \err{0.768}{0.257}{0.235} &  & 0.577 & \err{1.025}{0.348}{0.338} & \err{0.787}{0.036}{0.036} & 1.908 \\
\hline\hline
\end{tabular*}
\caption{Fitted parameters in the shear and bulk channels for all temperatures. We determine the confidence interval by taking the median of the bootstrap distribution as the central value, with the 34th percentiles on either side defining the lower and upper bounds.}
\label{tab:parameters}
\end{table*}
\begin{figure*}[bht]
\centerline{
\includegraphics[width=0.5\textwidth]{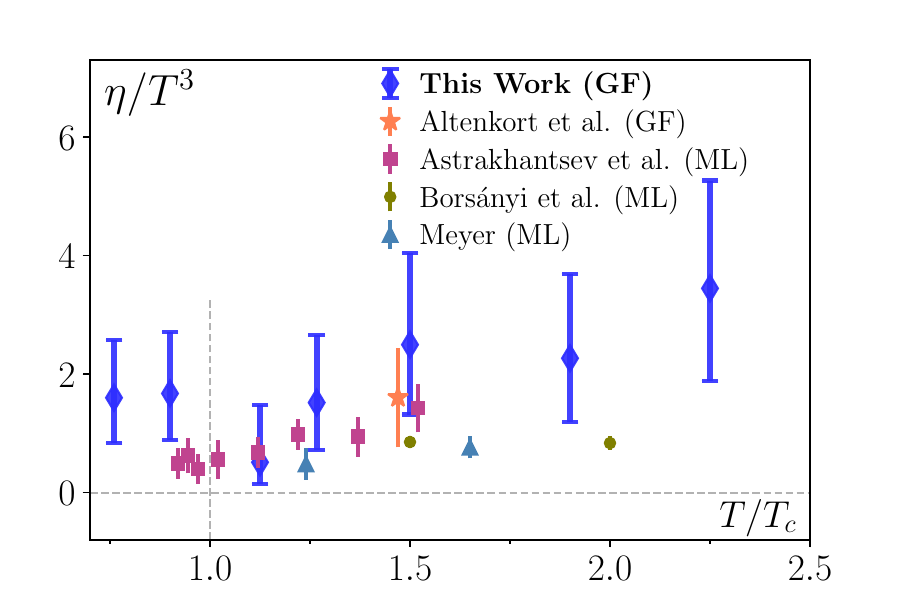}
\includegraphics[width=0.5\textwidth]{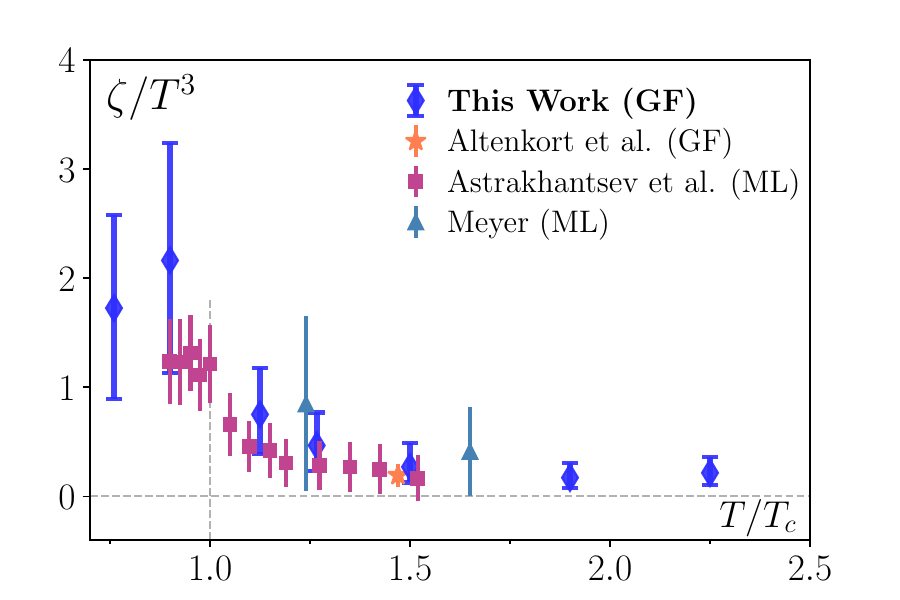}}
\caption{Comparison of the shear viscosity (left) and bulk viscosity (right) normalized by $T^3$. Lattice QCD results are categorized by algorithm: gradient flow is used in this work and in Ref.~\cite{Altenkort:2022yhb}, and the multilevel algorithm is employed by Astrakhantsev \textit{et al.}~\cite{Astrakhantsev:2017nrs,Astrakhantsev:2018oue}, Borsányi \textit{et al.}~\cite{Borsanyi:2018srz}, and Meyer~\cite{Meyer:2007ic,Meyer:2007dy}.}
\label{fig:vis_T3}
\end{figure*}
We find that the model provides a good description of our data in both channels across most temperatures, with the exception of the shear channel at $0.76\,T_c$ and $0.9\,T_c$.
These exceptions suggest that the spectral function below $T_c$ undergoes structural changes that are not captured by the perturbative UV form. 
To model this behavior while maintaining simplicity, we introduce a smoothing function
\begin{equation}
m(\omega/T) =\frac{1}{1+\exp\big{(}(\omega/T-\omega_0/T)/\Delta\big{)}} 
\label{eq:transition}
\end{equation}
which enforces a smooth crossover transition between the IR and UV regimes, 
\begin{equation}
\begin{split}
\frac{\rho^{\mathrm{cross}}_{\mathrm{model}}(\omega)}{\omega T^3}=&\frac{A}{T^3}\frac{C^2}{C^2+(\omega/T)^2}m(\omega/T)\\
&+B\frac{\rho^{\mathrm{pert}}(\omega)}{\omega T^3}\big{(}1-m(\omega/T)\big{)}\, . \\
\end{split}
\label{eq:modified_spf_model}
\end{equation}
In the absence of prior knowledge, we treat both the transition position $\omega_0/T$ and width $\Delta$ as free parameters. We find that $\omega_0/T$ is well constrained by fits to our data, while $\Delta$ values in the range [0.1, 10] all describe the data comparably well, yielding $\chi^2/{\rm d.o.f.}\in[0.202,0.922]$ for $0.76\,T_c$ and $\chi^2/{\rm d.o.f.}\in[0.325,1.192]$ for $0.9\,T_c$. However, for $\Delta\leq 0.5$, the extracted $\eta/T^3$ reaches a plateau---see Fig.~\ref{fig:0.76_eta_Delta} for an example obtained with $C=1$ at 0.76$\,T_c$, and it is similar for $C=5$ and $T=0.9\,T_c$. A previous lattice study~\cite{Astrakhantsev:2017nrs} using a similar model with higher data precision found $\Delta\sim0.25$ at $0.9\,T_c$, with values varying mildly between 0.19 and 0.43 across $T\in[0.9, 1.5]\,T_c$. These values lie within our plateau region. We therefore adopt $\Delta=0.25$ for our final estimates. With this crossover modification, including anomalous dimensions becomes unnecessary. We also adjust the shear channel switching scale $\mu_{\rm{shear}}^{\rm{swi}}$ to $11T$, which represents a typical value of $\omega_0$ from our fits. We have verified that this specific choice of $\mu_{\rm{shear}}^{\rm{swi}}$ has minimal impact on our results---using the original value of $2.146\pi T$ changes the obtained $\eta/T^3$ by less than 0.1\%.

The fitted correlators and spectral functions are shown in Fig.~\ref{fig:spf_shear_bulk}, using the shear channel at $0.76\,T_c$ and both channels at $1.5\,T_c$ as illustrative examples. Overall, the model describes the lattice data well. The fit parameters are summarized in Table~\ref{tab:parameters}, and the resulting viscosities normalized by $T^3$ are shown in Fig.~\ref{fig:vis_T3}. The central values in Fig.~\ref{fig:vis_T3} are taken as the average of the upper bounds obtained with $C=1$ and the lower bounds obtained with $C=5$. The error bars are obtained by adding the statistical uncertainties $\delta_{\rm stat}$ and systematic uncertainties $\delta_{\rm sys}$ in quadrature, with $\delta_{\rm sys}$ estimated as half of the differences between the results obtained with $C=1$ and $C=5$. Figure~\ref{fig:vis_T3} (left) shows that the values of $\eta/T^3$ obtained in this work and in earlier lattice calculations employing the multilevel algorithm~\cite{Meyer:2007ic,Astrakhantsev:2017nrs} exhibit a similar temperature dependence. Both Ref.~\cite{Astrakhantsev:2017nrs} and this work indicate that $\eta/T^3$ develops a dip in the vicinity of $T_c$. Below $T_c$, the results of Ref.~\cite{Astrakhantsev:2017nrs} and this work suggest that $\eta/T^3$ shows little temperature dependence, with values smaller than those near $T_c$. Above $T_c$, Refs.~\cite{Meyer:2007ic,Astrakhantsev:2017nrs} and this work consistently find that $\eta/T^3$ increases with temperature, and the results of Ref.~\cite{Astrakhantsev:2017nrs} are compatible with ours within $1\sigma$. In contrast, Ref.~\cite{Borsanyi:2018srz} reports that $\eta/T^3$ is largely insensitive to temperature in the range $1.5\,T_c$--$2.0\,T_c$. The tension among different lattice calculations observed in the left panel of Fig.~\ref{fig:vis_T3} is reduced in the right panel shown for $\zeta/T^3$. Both Ref.~\cite{Astrakhantsev:2018oue} and this work indicate that $\zeta/T^3$ develops a peak near $T_c$. Below $T_c$, these studies suggest that $\zeta/T^3$ exhibits little temperature dependence, with values tending to be peaked near $T_c$. Above $T_c$, all lattice calculations, including Refs.~\cite{Meyer:2007dy,Astrakhantsev:2018oue,Altenkort:2022yhb} and this work, find that $\zeta/T^3$ decreases with increasing temperature, with good agreement in magnitude among different studies.

\begin{table*}[t]
\centering
\renewcommand{\arraystretch}{1.5}
\begin{tabular}{c|c|c|c|c|c|c|c}\hline \hline
  & 0.76$\,T_c$ & 0.9$\,T_c$ & 1.125$\,T_c$ & 1.27$\,T_c$ & 1.5$\,T_c$ & 1.9$\,T_c$ & 2.25$\,T_c$ \\ \hline 
$\eta/s$ & \err{64.030}{39.079}{30.547} & \err{22.303}{13.852}{10.562} & \err{0.138}{0.260}{0.098} & \err{0.345}{0.260}{0.182} & \err{0.513}{0.319}{0.242} & \err{0.437}{0.275}{0.208} & \err{0.647}{0.342}{0.293} \\
$\zeta/s$ & \err{69.021}{34.131}{33.171} & \err{28.835}{14.325}{13.744} & \err{0.202}{0.116}{0.098} & \err{0.106}{0.069}{0.053} & \err{0.056}{0.045}{0.029} & \err{0.033}{0.026}{0.018} & \err{0.040}{0.028}{0.020} \\ \hline \hline
\end{tabular}
\caption{Viscosities normalized by the entropy density, obtained by dividing the $T^3$-normalized viscosities in Table~\ref{tab:parameters} by the entropy densities from Ref.~\cite{Giusti:2025fxu}, where uncertainties in the entropy density are neglected.}
\label{tab:viscosity-by-s}
\end{table*}

\begin{figure*}[htb]
    \centerline{
    \includegraphics[width=0.5\textwidth]{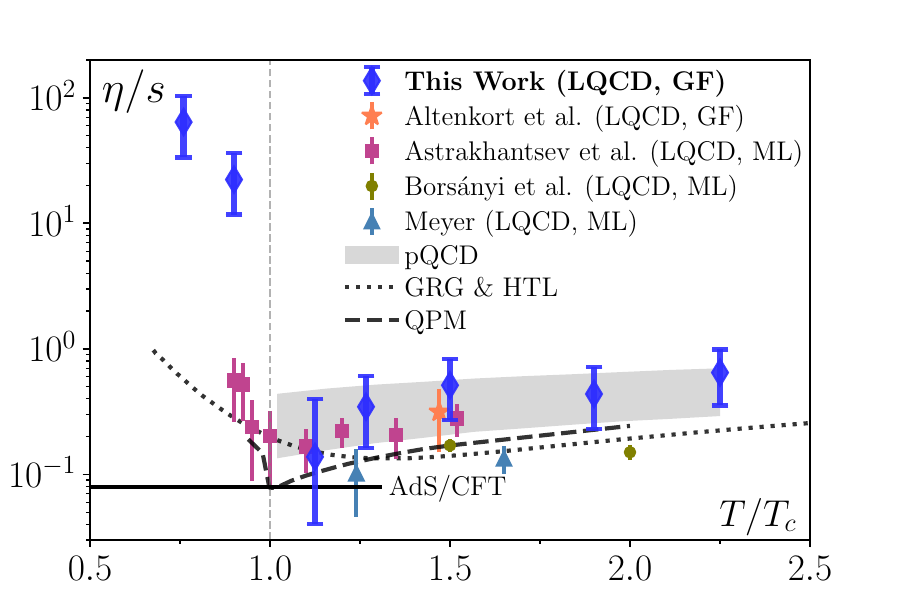}
    \includegraphics[width=0.5\textwidth]{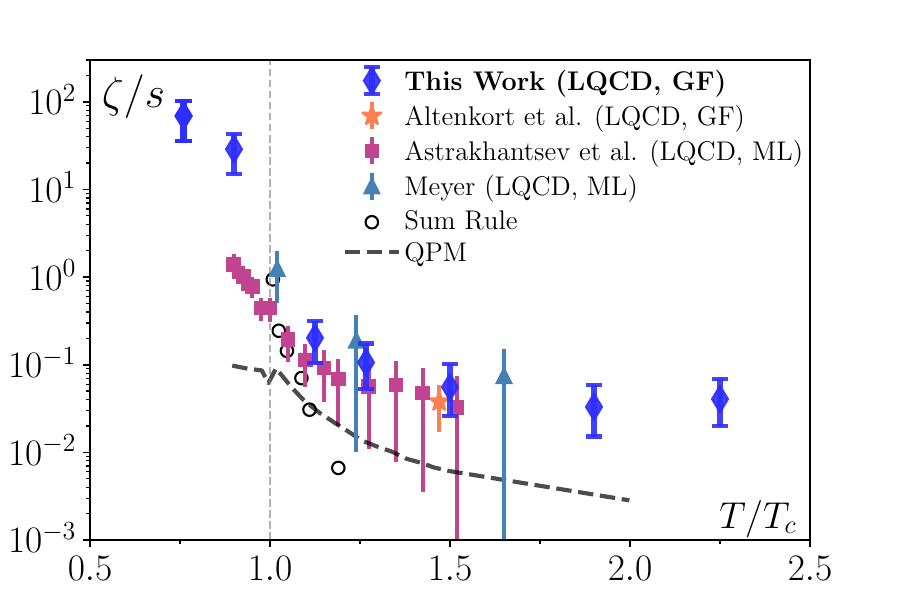}}
    \caption{Temperature dependence of $\eta/s$ (left) and $\zeta/s$ (right). Lattice QCD results are categorized by algorithm: gradient flow is used in this work and by Altenkort \textit{et al.}~\cite{Altenkort:2022yhb}, while the multilevel algorithm is employed by Astrakhantsev \textit{et al.}~\cite{Astrakhantsev:2017nrs,Astrakhantsev:2018oue}, Borsányi \textit{et al.}~\cite{Borsanyi:2018srz}, and Meyer~\cite{Meyer:2007ic,Meyer:2007dy}. For $\eta/s$, also shown are the NLO perturbative QCD calculation~\cite{Ghiglieri:2018dib}, an analytic fit employing glueball resonance gas for $T<T_c$ and HTL-resummed perturbation theory for $T>T_c$ (GRG and HTL)~\cite{Christiansen:2014ypa}, a quasiparticle model~\cite{Mykhaylova:2019wci}, and the AdS/CFT lower bound $1/(4\pi)$~\cite{Kovtun:2004de}. For $\zeta/s$, further comparisons include a sum rule analysis based on lattice QCD data (Sum Rule)~\cite{Kharzeev:2007wb} and a QPM result~\cite{Mykhaylova:2020pfk}.
    }
\label{fig:vis_all}
\end{figure*}

It is customary in the field to normalize viscosities by the entropy density $s$. We therefore also present comparisons of entropy-normalized viscosities from various studies in Fig.~\ref{fig:vis_all}. In our work, $s$ is obtained from an interpolation (or extrapolation) of the results in Ref.~\cite{Giusti:2025fxu}. The uncertainty in $s$ is neglected, since it is significantly smaller than that of the viscosities. The resulting entropy-normalized viscosities are listed in Table~\ref{tab:viscosity-by-s}. 

Our results for $\eta/s$ presented in the left panel of Fig.~\ref{fig:vis_all} show a dip near the phase transition temperature $T_c$, followed by a mild increase at $T>T_c$, approaching the AdS/CFT lower bound $1/(4\pi)$~\cite{Kovtun:2004de} near $1.125\,T_c$, in alignment with the NLO perturbation theory~\cite{Ghiglieri:2018dib} (perturbation QCD, ``pQCD") at higher temperatures.
In comparison, an analytic fit employing glueball resonance gas (GRG) for $T<T_c$ and HTL-resummed perturbation theory for $T>T_c$ (``GRG and HTL'')~\cite{Christiansen:2014ypa}, the NLO perturbation theory (``pQCD"), and the quasiparticle model (``QPM")~\cite{Mykhaylova:2019wci}, all predict a similar dip and mild rise, consistent with the trend seen in our results. Two lattice calculations employing the ML algorithm~\cite{Astrakhantsev:2017nrs,Meyer:2007ic} (``Astrakhantsev \textit{et al.}” and ``Meyer”) reproduce the mild increase above $T_c$ observed in our study in a narrower temperature range. Another lattice  study using ML~\cite{Borsanyi:2018srz} (``Borsányi \textit{et al.}") reports little or no temperature dependence. 

For $\zeta/s$ presented in the right panel of Fig.~\ref{fig:vis_all}, our results exhibit a decreasing trend with temperature from below $T_c$ to above $T_c$. Other calculations—including the ``QPM"~\cite{Mykhaylova:2020pfk}, sum rule analyses combined with lattice data (``Sum Rule")~\cite{Kharzeev:2007wb}, and ML-based lattice studies~\cite{Astrakhantsev:2018oue,Meyer:2007dy} (``Astrakhantsev \textit{et al.}" and ``Meyer")—show a similar trend for $T>T_c$ but report smaller values below $T_c$ when normalized by the entropy density.

Figure~\ref{fig:vis_all} shows that, for $T<T_c$, both $\eta/s$ and $\zeta/s$ obtained in this work are significantly larger than the results from the analytic fit~\cite{Christiansen:2014ypa}, the QPM~\cite{Mykhaylova:2019wci}, and the lattice determinations of Astrakhantsev \textit{et al.}~\cite{Astrakhantsev:2017nrs,Astrakhantsev:2018oue}. 
However, when the viscosities are expressed in units of $T^3$, the differences between the lattice results of Refs.~\cite{Astrakhantsev:2017nrs,Astrakhantsev:2018oue} and this work are substantially reduced, as shown in Fig.~\ref{fig:vis_T3}. 
This indicates that the dominant source of the discrepancy originates from the determination of the entropy density $s$. In the present study, $s$ is taken from the high-precision, continuum-extrapolated results of Ref.~\cite{Giusti:2025fxu} obtained using shifted boundary conditions, whereas Refs.~\cite{Astrakhantsev:2017nrs,Astrakhantsev:2018oue}, as well as GRG and HTL~\cite{Christiansen:2014ypa} and QPM~\cite{Mykhaylova:2019wci}, rely on entropy densities computed using the integral method at finite lattice spacing~\cite{Engels:1999tk,Borsanyi:2012ve}, which becomes less reliable below $T_c$ due to the smallness of pressure and plaquette differences.

To complement the quantitative comparison presented above, it is instructive to briefly contrast our analysis with previous lattice extractions of the shear and bulk viscosities from Euclidean energy-momentum tensor correlators. The differences most relevant for such a comparison are (i) how the dominant modeling systematics associated with the low-frequency transport peak are quantified, (ii) the control over lattice spacing effects and the statistical precision of the correlators, and (iii) the temperature range covered.

Most importantly, our analysis treats the transport peak width as an explicit source of systematic uncertainty in the spectral Ans\"atze by scanning a physically motivated range set by thermal scales. 
As summarized in Table~\ref{tab:parameters}, at the upper end of this range, $C=5$ (which corresponds to the lower bound of the extracted viscosity), the uncertainty budget is typically dominated by the systematic component. Quantitatively, the systematic contribution accounts for $\gtrsim \mathcal{O}(70\%)$ of the total uncertainty, defined as $\delta_{\rm sys}^2/(\delta_{\rm sys}^2+\delta_{\rm stat}^2)$, in both the shear and bulk channels, except for the shear viscosity at $T=1.125\,T_c$.
Consequently, the total error bars, defined as $\sqrt{\delta_{\rm sys}^2+\delta_{\rm stat}^2}$, are comparatively larger than those reported in determinations employing the multilevel algorithm~\cite{Meyer:2007ic,Meyer:2007dy,Astrakhantsev:2017nrs,Astrakhantsev:2018oue,Borsanyi:2018srz}. 
This is evident both for the $T^3$-normalized viscosities shown in Fig.~\ref{fig:vis_T3} and for the entropy-normalized results in Fig.~\ref{fig:vis_all}. 
The multilevel-based determinations typically rely on the Backus-Gilbert method or on model fits in which the transport peak is taken to be nearly flat.
Equivalently, the intrinsic curvature at low frequency is either assumed negligible or is washed out by averaging over a finite low-frequency window.
As a result, their reported central values align more closely with our results obtained for a broader transport peak, corresponding to $C=5$. 
Notably, an earlier study, Ref.~\cite{Altenkort:2022yhb} [labeled ``Altenkort \textit{et al.} (GF)” in Fig.~\ref{fig:vis_all}], which employs a spectral reconstruction strategy similar to that used here, yields results that are consistent with ours within uncertainties.

Complementing this treatment of the dominant spectral modeling systematics, we also improve control over discretization effects, which in practice are governed by the lattice temporal extent $N_\tau$ (or equivalently at fixed $T$, by the lattice spacing) in the calculations compared below.
To make the comparison concrete, we consider a representative moderate temperature: 
$1.65\,T_c$ for Refs.~\cite{Meyer:2007ic,Meyer:2007dy} and $1.5\,T_c$ for Refs.~\cite{Astrakhantsev:2017nrs,Astrakhantsev:2018oue,Borsanyi:2018srz} as well as this work. Whenever continuum extrapolations are available, we quote the finest lattice (largest $N_\tau$) used in the analysis. 
At these temperatures, the largest temporal extents in Ref.~\cite{Meyer:2007ic}, Ref.~\cite{Meyer:2007dy}, Refs.~\cite{Astrakhantsev:2017nrs,Astrakhantsev:2018oue}, Ref.~\cite{Borsanyi:2018srz}, and this work are $N_\tau=8$, $12$, $16$, $20$, and $36$, respectively, corresponding to finest lattice spacings $a=0.047$, $0.032$, $0.025$, $0.021$, and $0.012~\mathrm{fm}$.
Taken together, the availability of multiple finer and larger lattices in the present study enables a more reliable extrapolation to the continuum limit. Alongside this improved control of discretization effects, the combination of gradient flow and blocking methods yields good statistical precision. In particular, the relative uncertainties of the correlators at $\tau T=0.5$ in both the shear and bulk channels lie in the range of 3\%--12\% across all temperatures. This level of precision is comparable to that achieved in studies using the multilevel algorithm: Ref.~\cite{Meyer:2007ic} reports uncertainties of 5\% and 6\% at $1.24\,T_c$ and $1.65\,T_c$, while Refs.~\cite{Astrakhantsev:2017nrs,Astrakhantsev:2018oue} and Ref.~\cite{Borsanyi:2018srz} report uncertainties below 2\%--3\% and 2\%, respectively. 
It is worth noting that the statistical precision achieved in this work relies on a comparatively modest ensemble size of about $5\times10^3$ configurations. 
By contrast, multilevel-based calculations often rely on substantially larger statistics, for example, Ref.~\cite{Borsanyi:2018srz} uses ensembles of order millions of configurations.
The combined control over both systematic and statistical uncertainties thus provides more stringent constraints on the spectral modeling.

Finally, the temperature range studied here extends from $0.76\,T_c$ up to
$2.25\,T_c$, substantially broader than the widest range $0.9\,T_c$--$1.5\,T_c$
explored in the previous lattice studies.
We find that for $T<T_c$, down to $0.76\,T_c$, both $\eta/T^3$ and $\zeta/T^3$ remain largely insensitive to temperature, reinforcing the behavior observed in Refs.~\cite{Astrakhantsev:2017nrs,Astrakhantsev:2018oue} previously established only down to $0.9\,T_c$. 
Extending the analysis to higher temperatures is essential for making contact with perturbation theory, as perturbative calculations become increasingly reliable in this regime. 
Indeed, we observe that the agreement between the perturbative calculations (``pQCD”) and this work improves with increasing temperature, thereby justifying the use of NLO spectral functions in the spectral modeling.

\section{Conclusion}
\label{sec:conclusion}
We have determined the temperature dependence of the shear and bulk viscosities of SU(3) Yang-Mills theory in the range  $0.76\,T_c \le T \le 2.25\,T_c$. 
Our strategy is based on high-precision Euclidean correlators of the energy-momentum tensor, obtained with the gradient flow-defined EMT combined with the blocking method.
At each temperature we perform calculations on three fine and large lattices (up to $L=3.31~\mathrm{fm}$ and down to $a=0.01164~\mathrm{fm}$), enabling controlled continuum extrapolations of the correlators. Shear and bulk viscosities are extracted by fitting correlators to models constructed using the NLO perturbative spectral functions, which provide currently the best known ultraviolet input. The dominant systematic uncertainty in the extraction of viscosities originates from  the (\textit{a priori} unknown) transport peak width. We quantify this by bracketing the width within a physically motivated range set by thermal scales (our choices of $C=1$ and $C=5$). The fitted parameters and resulting $T^3$-normalized viscosities are summarized in Table~\ref{tab:parameters} and Fig.~\ref{fig:vis_T3}, while the entropy-normalized results are given in Table~\ref{tab:viscosity-by-s} and compared with other determinations in Fig.~\ref{fig:vis_all}.

Our main findings can be summarized as follows. In units of $T^3$, $\eta/T^3$ exhibits a dip in the vicinity of $T_c$ and increases with temperature for $T>T_c$, whereas below $T_c$ (down to $0.76\,T_c$) it shows only a weak temperature dependence; see Fig.~\ref{fig:vis_T3} (left). For the bulk channel, $\zeta/T^3$ develops a peak near $T_c$ and decreases in the deconfined phase, with again a mild temperature dependence below $T_c$, see Fig.~\ref{fig:vis_T3} (right).
Normalizing by the continuum-extrapolated entropy density from Ref.~\cite{Giusti:2025fxu}, we find that $\eta/s$ reaches a minimum in the transition region and then rises mildly for $T>T_c$, while $\zeta/s$ decreases with increasing temperature across the deconfined regime, see Fig.~\ref{fig:vis_all} and Table~\ref{tab:viscosity-by-s}. At high temperature, our $\eta/s$ results are consistent with the NLO perturbative prediction~\cite{Ghiglieri:2018dib} and approach it increasingly well as $T$ increases, see Fig.~\ref{fig:vis_all}. Near the transition region, the minimum values of $\eta/s$ are close to the AdS/CFT lower bound $1/(4\pi)$~\cite{Kovtun:2004de}, providing a useful benchmark for the strongly coupled regime.


\section*{DATA AVAILABILITY}
The data that support the findings of this article are openly available~\cite{ding_2026_18976353}, embargo periods may apply.

\section*{ACKNOWLEDGMENTS}
We thank Guy D. Moore for helpful discussions. This work is supported partly by the National Natural Science Foundation of China under Grants No.~12325508, No.~12293064, and No.~12293060 , as well as the National Key Research and Development Program of China under Contract No.~2022YFA1604900 and the Fundamental Research Funds for the Central Universities, Central China Normal University, under Grants No.~30101250314 and No.~30106250152. The numerical simulations have been performed on the GPU cluster in the Nuclear Science Computing Center at
Central China Normal University (NSC3) using SIMULATeQCD suite~\cite{Altenkort:2021cvg,mazur2021,HotQCD:2023ghu}.

\begin{widetext}
\section*{Appendix: JOINT CONTINUUM EXTRAPOLATION}
\appendix


\begin{figure*}[htb]
    \centerline{
    \includegraphics[width=0.5\textwidth]{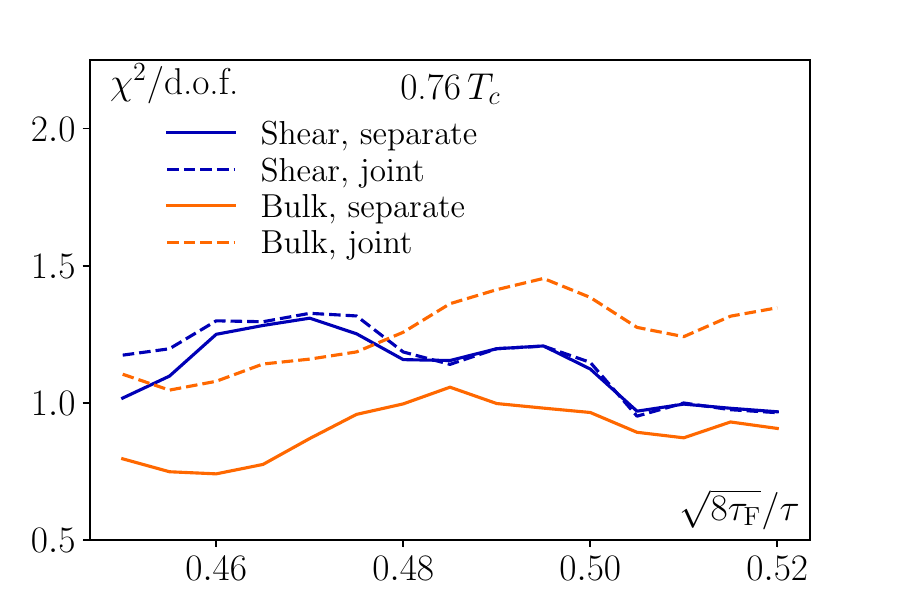}
    \includegraphics[width=0.5\textwidth]{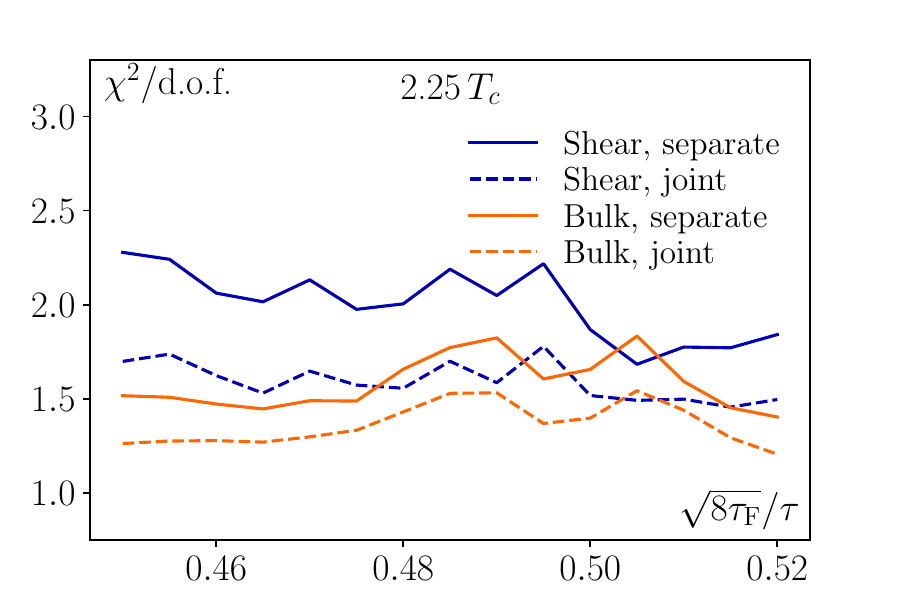}}
    \caption{Left: comparison of the $\chi^2$/d.o.f. from the joint and separate continuum extrapolations for both shear and bulk channels at $0.76\,T_c$. Right: same as the left panel but for $2.25\,T_c$. For the separate fits the $\chi^2$/d.o.f. is the averaged one over all usable $\tau T$.
    \label{fig:compare-chisq}
}
\end{figure*}
In Fig.~\ref{fig:compare-chisq}, we compare the $\chi^2$/d.o.f. from the joint and separate fits for both shear and bulk channels, using the lowest temperature $0.76\,T_c$ and highest temperature $2.25\,T_c$ as an example. The results demonstrate that joint fits generally provide a better description of our lattice data, achieving smaller $\chi^2$/d.o.f. values compared to separate fits. We note that this trend persists across other temperatures, though not shown here.

\begin{figure*}[t]
    \centerline{
    \includegraphics[width=0.5\textwidth]{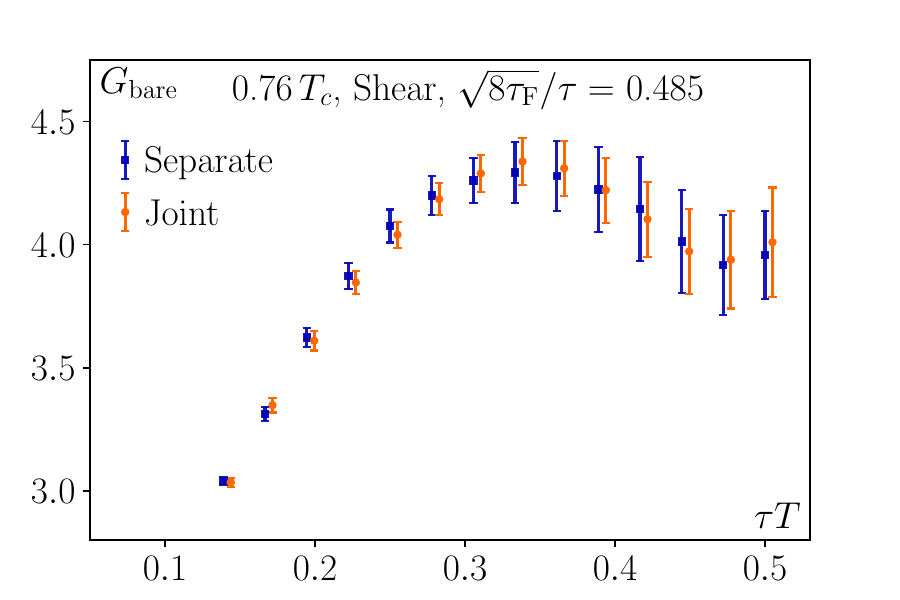}
    \includegraphics[width=0.5\textwidth]{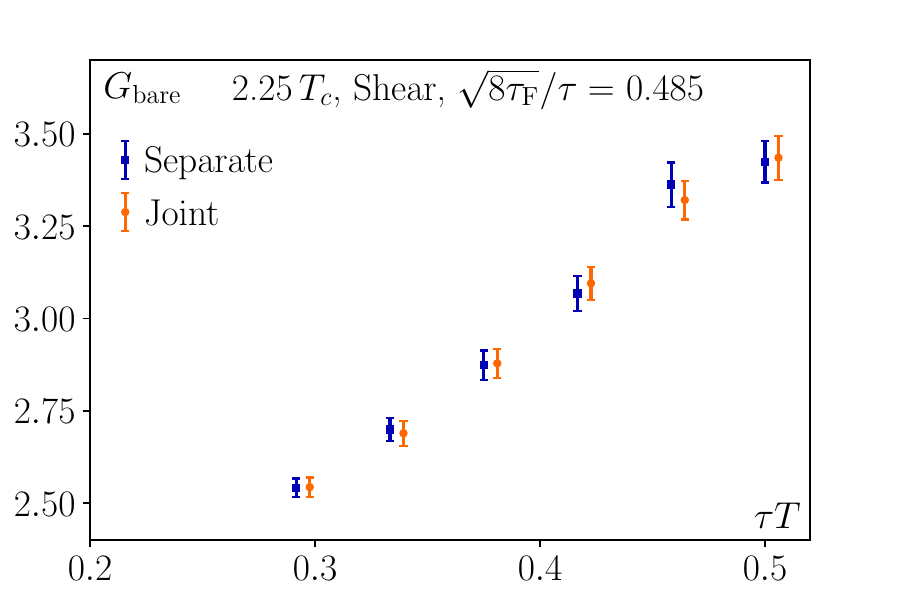}}
    \centerline{
    \includegraphics[width=0.5\textwidth]{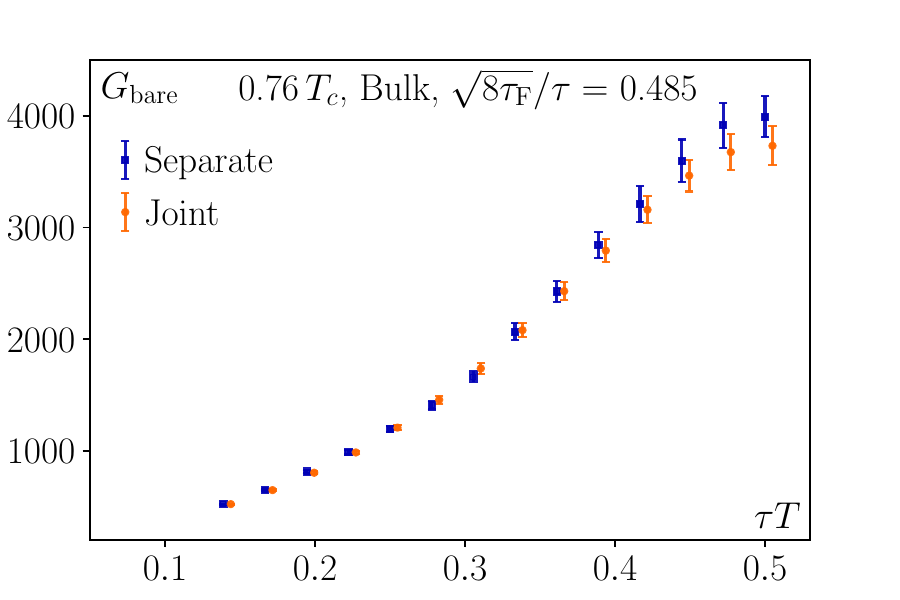}
    \includegraphics[width=0.5\textwidth]{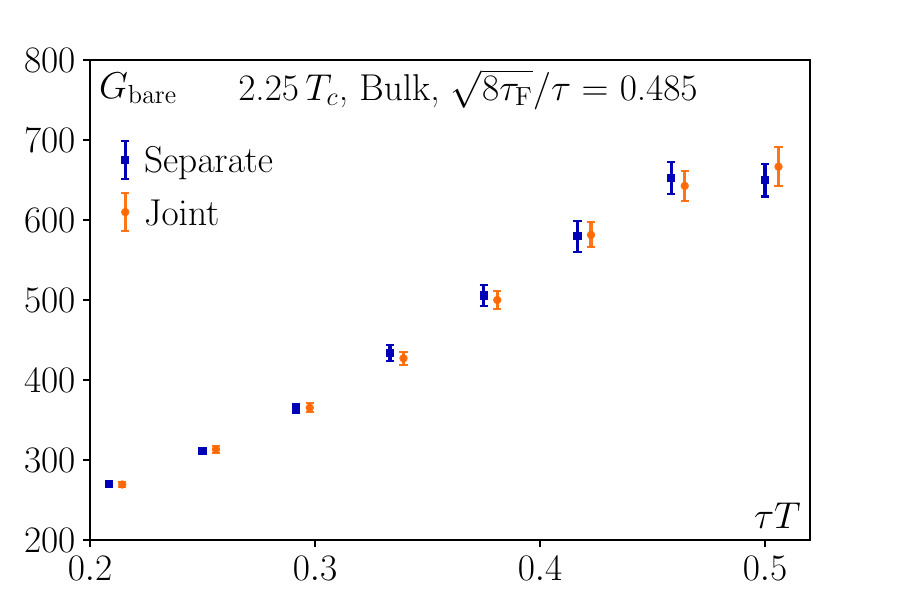}}
    \caption{Top left: comparison of the continuum-extrapolated correlators from the joint and separate fits in the shear channel at $T=0.76\,T_c$ and $\sqrt{8\tauf}/\tau=0.485$.
    Top right: same as the left panel but for $T=2.25\,T_c$. 
    Bottom: same as the top panels but for the bulk channel. The data points have been slightly shifted horizontally for better visibility.
}
\label{fig:compare-corr}
\end{figure*}

Figure~\ref{fig:compare-corr} compares continuum-extrapolated correlators from joint and separate fits for both shear and bulk channels at temperatures $T=0.76\,T_c$ and $2.25\,T_c$, evaluated at $\sqrt{8\tauf}/\tau=0.485$. The results demonstrate agreement between joint and separate fits within 1$\sigma$ statistical errors across all cases.
\end{widetext}

\FloatBarrier
\bibliographystyle{apsrev4-1}
\bibliography{paper}

\end{document}